# IoT-Driven Building Energy Management Systems (BEMS) for Net Zero Energy Buildings: Concept, Integration and Future Directions

Haizum Hanim Ab Halim[1], Dalila Alias[1], Akmal Zaini Arsad[1], Lewis Tee Jen Looi[1], Rosdiadee Nordin[2,3], Denny Ng Kok Sum[1]

1. Faculty of Engineering and Technology, Sunway University, No 5, Jalan Universiti, Bandar Sunway, 47500, Selangor, Malaysia
2. Future Cities Research Institute, Faculty of Engineering and Technology, Sunway University, No5, Jalan Universiti, Bandar Sunway, 47500, Selangor, Malaysia
3. Future Cities Research Institute, Lancaster University, Lancaster, LA1 4YW, United Kingdom

**Abstract:** Construction and operating of buildings is one of the major contributors to global greenhouse emissions. With the inefficient usage of energy due to human behavior and manual operation, the energy consumption of buildings is further increased. These challenges highlight the need for improved Building Energy Management Systems (BEMS) integrated with Internet of Things (IoT) and data driven intelligence to enhance energy-efficiency in a building and contribute to Net-Zero Energy Buildings (NZEB) targets. This paper offers four keys contributions: i) a systematic review of IoT enabled BEMS including components, network architecture and functional capabilities, ii) an evaluation of real-world BEMS datasets to support Artificial Intelligence (AI) based predictive control, iii) an analysis of integration challenges related to interoperability, smart grids and net-zero energy strategies, and iv) a case study highlighting global best practices, performances outcomes, and lesson learned for scaling advanced BEMS solutions.

___________________________________
Corresponding Author:
Rosdiadee Nordin, rosdiadeen@sunway.edu.my
Other Authors:
Haizum Hanim Ab Halim, haizumh@sunway.edu.my
Dalila Alias, dalilaa@sunway.edu.my
Akmal Zaini Arsad, akmalzaini@sunway.edu.my
Lewis Tee Jen Looi, lewist@sunway.edu.my
Denny Ng Kok Sum, dennyng@sunway.edu.my

## 0  Introduction

Environmental sustainability is directly threatened by carbon emissions from the building and energy industries has hasten climate change. Persistent energy wastage and low energy efficiency may raise resource consumption and operating expenses due to increasing carbon emissions and impeding sustainable growth.

One of the major contributors to greenhouse gases (GHG) emissions is energy usage and wastage in buildings, which accounts for nearly 40% of global energy consumption and a significant portion of carbon emissions [1]. Meanwhile, embodied carbon emissions from materials and construction processes contributes another 10% of global carbon emissions attributable to buildings [2]. Energy wastage in buildings primarily results from inefficient heating, ventilation, air conditioning (HVAC) systems, poor insulation, outdated appliances, and excessive lighting usage, all of which increase energy demand and carbon emissions. As buildings rely heavily on electricity generated from fossil fuels, unnecessary energy consumption significantly amplifies greenhouse gas emissions, making energy efficiency improvements crucial for reducing the environmental impact of the built environment. Inefficient energy usage in HVAC, lighting, and appliances leads to excessive electricity demand, further increasing the reliance on fossil fuel-based energy production and exacerbating environmental degradation [3].

The surge in energy costs has placed financial strain on businesses and industries, leading them to look for more effective energy management strategies to retain profitability and operational stability. The crisis worsens after Russia's invasion of Ukraine in 2022, causing disruptions in global energy markets that contributed to the sharp rise in energy costs, driven by fossil fuel price vitality, geopolitical instability, economic recovery and inflation, which has highlighted the urgent need for energy-efficient solutions and the transition to renewable energy sources [4]. Furthermore, the post-pandemic economic rebound has driven up global energy demand, intensifying competition for limited resources and straining existing energy infrastructure.

The shift toward renewable energy, including solar and wind power, is driven by the urgent need to mitigate climate change and reduce dependence on fossil fuels, which contribute to greenhouse gas emissions [5]. However, integrating these renewable sources into existing power grids presents challenges such as intermittency, storage limitations, and demand fluctuations, requiring intelligent energy management solutions. Building Energy and Management Systems (BEMS) play a crucial role in addressing these issues by monitoring, analyzing, and optimizing energy consumption in buildings to ensure efficient use of renewable energy [6]. By leveraging IoT, AI, and automation, BEMS can enhance energy efficiency, reduce operational costs, and support the transition to smart, sustainable buildings to achieve net zero emissions. As a result, the widespread adoption of BEMS not only helps organizations achieve cost savings and regulatory compliance but also contributes significantly to global sustainability goals and carbon neutrality efforts [6].

To address the challenges of excessive energy consumption and waste in buildings, the implementation of BEMS has emerged as a crucial solution for optimizing energy usage and improving sustainability. Recognizing the importance of energy-efficient buildings, governments and regulatory bodies worldwide have introduced policies and standards to promote sustainable practices. The United States has implemented the Energy Policy Act, which encourages the adoption of smart energy management technologies by providing incentives and regulatory frameworks to enhance energy efficiency in buildings [3]. These initiatives underscore the growing global commitment to integrating intelligent energy management solutions like BEMS to mitigate the environmental impact of buildings and support long-term sustainability.

In Malaysia, National Energy Policy places strong emphasis on the integration of energy-efficient appliances, smart grids, and renewable energy sources to enhance building sustainability and reduce overall energy consumption [7]. This policy aims to promote energy efficiency by encouraging the adoption of advanced technologies such as smart meters, automated energy management systems, and intelligent lighting and HVAC controls, which help optimize energy usage and minimize waste. Additionally, the implementation of smart grids plays a crucial role in improving energy distribution by enabling real-time monitoring, demand-response strategies, and better integration of renewable energy sources such as solar and wind power. By transitioning towards cleaner and more efficient energy systems, Malaysia aligns with global sustainability efforts, including the United Nations Sustainable Development Goals (SDG 7 – Affordable and Clean Energy, and SDG 13 – Climate Action), which focus on ensuring universal access to reliable and sustainable energy while combating climate change [5]. However, despite these efforts, challenges such as high implementation costs, regulatory barriers, and the need for greater public awareness continue to hinder widespread adoption. Addressing these challenges through government incentives, policy refinements, and increased research and development investments will be critical in driving Malaysia's transition towards a more energy-efficient and environmentally sustainable building sector. As part of this transition, renewable energy sources such as solar and wind power are increasingly being integrated into BEMS to reduce dependence on fossil fuels and lower carbon emissions [8].

## 1 Background and Fundamentals

BEMS is a system that monitors and controls a building's energy use to reduce emissions, designed to be efficient, scalable, reliable, and flexible, with the ability to sense its environment and make autonomous decisions [9-10]. BEMS offer significant benefits by enhancing energy efficiency, reducing costs, and improving building performance. By integrating monitoring, analysis, and control systems, BEMS help organizations optimize HVAC, lighting, and other energy-consuming processes, leading to lower operational expenses and better resource utilization [11].

The application of BEMS extends to commercial, residential, and industrial buildings, where it ensures sustainability by minimizing energy waste and reducing carbon emissions. BEMS leverages advanced technologies such as real-time monitoring, AI and the IoT to analyze energy consumption patterns, predict demand, and automate system adjustments, ultimately enhancing efficiency, lowering operational costs, and reducing environmental impacts [12]. Additionally, advanced BEMS technologies integrate with smart systems like Building Information Modelling (BIM), AI, and IoT, allowing for real-time energy tracking, predictive maintenance, and automated control. Overall, BEMS not only enhance sustainability and regulatory compliance but also improve occupant comfort and operational efficiency [11, 13].

1.1 BEMS Components and Implementation

BEMS is supported by a multi-layered technical architecture integrating with hardware, software and communication infrastructure to ensure an affective energy monitoring and building automation. The components and implementation of BEMS include sensors, actuators, controllers, network communication, data management systems, and AI optimization.

The core components in BEMS are sensors as they initiate the BEMS process. Sensors are devices that can measure parameters, monitor environmental conditions and offer real-time data such as temperature, humidity, air quality, occupancy and energy consumption [14]. Sensors will measure and collect data for each parameter before sending it to a storage facility for further processing and decision making. Therefore, sensor selection must be selective and precise to ensure the data accuracy which improves system reliability, control efficacy and system performance. The sensor data is also used to forecast energy performance for early fault diagnosis.

Data acquired by sensors is subsequently sent to a controller, which makes decisions for the system to improve efficiency and comfort. These devices regulate and control the building's numerous systems, for example HVAC, lighting, renewable energy and energy usage [15–16]. The system's intelligence is determined by how the controller controls the system using the data provided. In addition to the controller, actuators such as valves were added in HVAC system in BEMS, which are responsible for the dynamic control of chilled water system to achieve specific temperature regulation and maximum energy efficiency of the system [17]. Actuators transform controller commands into a physical action, and without them, a smart BEMS is only a monitoring system.

Once the data has been collected and is ready to be sent to storage for further processing, reliable communication is needed to ensure the data is transferred securely and robustly. BEMS uses a variety of network communication protocols based on layer, for example RS485 connects sensors, meters and equipment such as lighting and HVAC component at physical layer, field level protocol like Modbus and BACnet/IP integrate with smart energy meter, photovoltaic systems, and battery storage system, network and supervisory layer protocol like Ethernet, WiFi and at application or IoT layer protocol like Message Queue Telemetry Transport (MQTT) and CoAP [18-19]. The communication layer in BEMS enables secure and interoperable data flow from field devices to centralized and cloud-based management platforms.

The acquired data are then transmitted to Data Management System (DMS) where they are collected, stored and analyzed. These systems enable data preprocessing, validation, normalization and secure storage to ensure data integrity and interoperability across heterogonous subsystems. Moreover, DMS supports advanced analytics, model-based simulation, and ML application for fault detection, performance optimization and predictive energy management. Figure 1 shows components in BEMS.

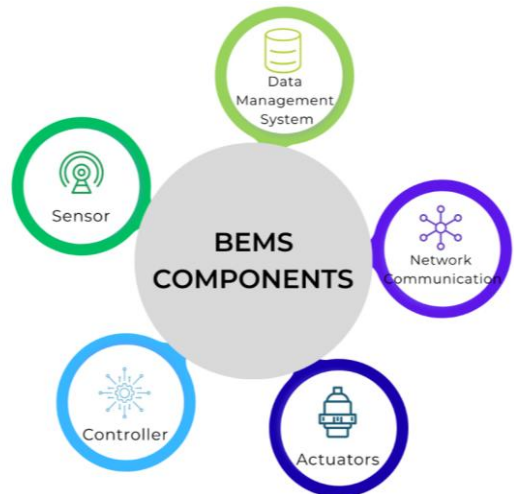

Fig. 1   BEMS components

1.2 BEMS Function and Features

Each component of BEMS has played an important role in improving energy efficiency in building operations to meet sustainability and cost reduction goals. Core BEMS's functions and features include energy monitoring, control and automation, energy management and optimization, maintenance support and system integration.

The main function of BEMS is tracking energy consumption, power monitoring, recording historical performance and performance tracking based on real-time data collected from sensors and actuators. Smart metering and monitoring systems are used to monitor features which are then shown on the BEMS dashboard. Monitoring features are performed by using smart metering and monitoring system and displaying in BEMS dashboard.

BEMS are also able to regulate building systems by control and automation function, such as HVAC system, smart lighting, plug load monitoring and control and access control. This function is conducted to reduce energy usage, minimize energy wastage while maintaining comfort and security. HVAC system maintaining comfort room temperature, meanwhile maximizing the energy efficiency [20], and smart plugs control energy usage by preventing unnecessary consumption [21].

Energy management and optimization are enabled through the Energy Management System (EMS) to provide building energy consumption analysis across multiple subsystems. The EMS applies optimization strategies such as load scheduling, real-time energy optimization, predictive maintenance, peak demand management, coordination of HVAC, fault detection and data security and interoperability [6].

Moreover, BEMS is equipped with system automation and maintenance that support fault detection and diagnostic (FDD) and predictive maintenance through data analytics and AI optimization [22]. FDD may detect early faults to avoid any system disruption and increasing energy efficiency. Even though BEMS concerns system operational and energy focused, safety and security were implemented to protect the building systems. CCTV with integration of access control uses smart authentication and real-time monitoring to secure entry points of the building [23]. Figure 2 shows the illustration of BEMS features and applications.

By enabling real-time data collection, automated control, and performance visualization, BEMS not only reduces wastage and operational costs but also improves occupant comfort and supports environmental objectives. Consequently, BEMS has emerged as a vital tool in modern smart buildings and is a key enabler of the global shift toward energy-efficient infrastructure and low carbon.

## 2.0 Net Zero Energy Building Concepts and Strategies

The building sector is a major contributor to global energy consumption and associated with GHG emissions, motivating the development of high energy performance building paradigms aimed at reducing operational energy demand. The concept of net-zero energy buildings (NZEBs) has emerged as a strategy through collective effort to reduce building-sector energy consumption and enhance sustainability of the built environment.

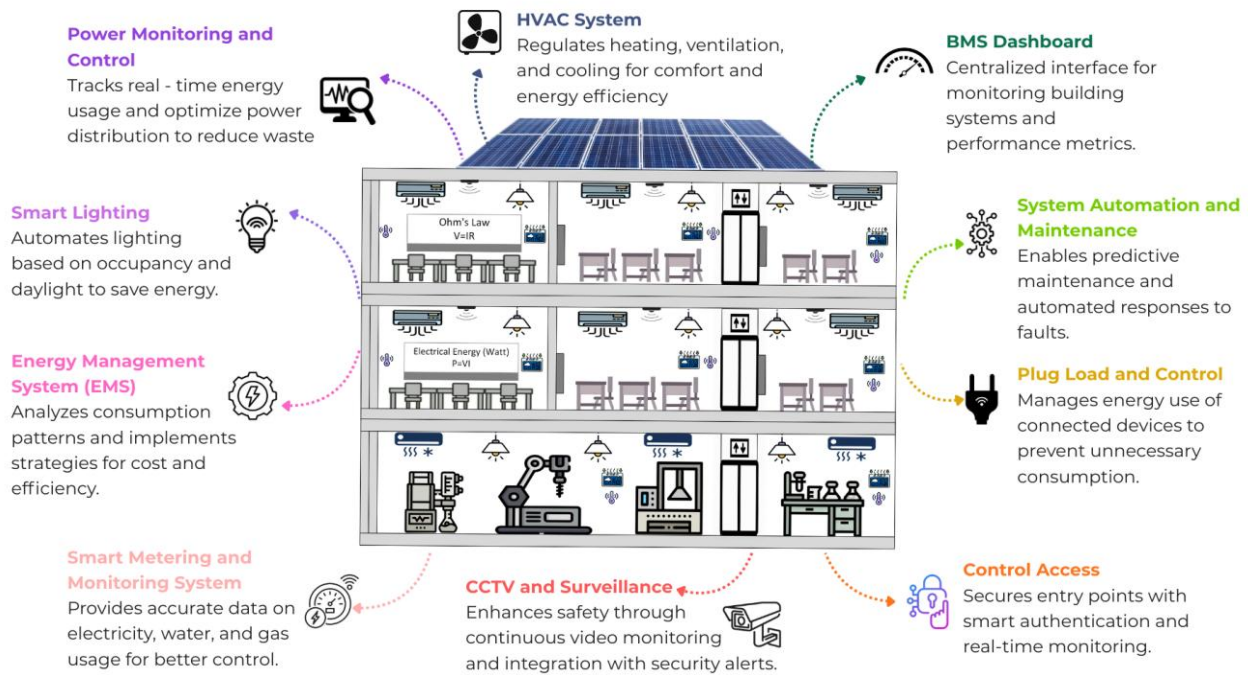

Fig. 2 BEMS features and applications

Early efforts in building energy performance focused on incremental energy efficiency improvements. Over time, this approach evolved towards outcome-based performance targets, culminating in the NZEB concept defined by a building's ability to offset its operational energy demand through renewable energy production, typically over a period of one year [24-25].

The NZEB concept gained institutional support through government agencies, professional bodies, and research organisations. The United States Department of Energy (DOE) articulated long term targets for achieving net zero energy performance across building sectors, and the American Society of Heating Refrigerating and Air-Conditioning Engineers (ASHRAE) highlighted the role of NZEB in future energy systems including integrated building, grids, and distributed energy resources [24, 26].

Despite widespread adoption of the term NZEB, there is no single universally standardized definition of NZEBs, with variations persisting across regions, standards, and literature [27]. [25] proposed a classification framework that distinguishes NZEBs based on the energy accounting boundary and performance metric, including "Net zero site energy", "Net zero source energy", "Net zero energy emission", and "Net zero energy cost". As each definition carries different emphasis on system boundaries and performance priorities, the selection of definition has a significant impact on subsequent building design, technology selection, and performance evaluation.

Among these, source-based and site-based energy definitions are most frequently adopted in literature and building energy practices with measurable operational flows. Net zero source energy building considers upstream energy losses associated with fuel extraction, power generation, transmission, and requires site-to-source conversion factors to reflect primary (source) energy usage. This provides a more comprehensive system level perspective at additional uncertainty from regional variability due to conversion factors and energy mixes. A net zero site energy building benefits from simplicity and direct verifiability through on-site metering, making it an attractive design for performance evaluation and tracking. A Net zero site energy building is defined as a building that produces the same amount of, or more, renewable energy as it consumes on an annual basis, with both generation and consumption being accounted for at the building site [24-25].

Besides energy accounting metrics, [28] proposed a widely referenced classification system that categorizes NZEBs based on the location and nature of renewable energy supply. This framework defines four NZEB classes (A–D), ranging from buildings that rely exclusively on renewable energy generated within the building footprint to those that depend on off-site renewable energy procurement. This provides a structured approach for comparing NZEB performance across projects with different physical and contextual constraints and is applicable to individual buildings as well as campus-scale or community-scale developments.

This typology also highlights the importance of prioritizing energy efficiency first, followed by demand reduction and demand optimization. After which, the demand is met by on-site renewable generation before resorting to off-site solutions. This approach emphasizes the importance of Net zero building design, system selection, and operational control in achieving Net zero enegy performance. Therefore, the need for effective monitoring and demand management strategies is evident.

Therefore, NZEBs could be understood as dynamic energy systems whose performance depends on interactions among building designs, grid connectivity, renewable energy availability, as well as occupant requirements and behavior. This perspective reinforces the need for advanced building energy management strategies, enabled by modern digital technologies and intelligent control systems as discussed in subsequent sections.

2.1 Design Strategies

Achieving net zero energy performance relies on minimizing building energy demand prior to supplying through clean, renewable energy generation. Design strategies for NZEBs therefore emphasize a hierarchical approach in which passive designs are first prioritized to reduce intrinsic loads, followed by active systems that efficiently serve the remaining demand [29-30]. This section reviews key NZEBs passive and active design strategies that function as enablers for a building to achieve net zero energy performance.

2.1.1 Passive Design Strategies

Passive design strategies aim to reduce heating, cooling, and lighting demand through architectural and lighting demand through architectural and envelope-level decisions that exploit local climate conditions. These strategies are particularly critical for NZEBs, as they provide sustainable, low-maintenance performance benefits and reduce long-term reliance on mechanical systems [30]. Common strategies include building orientation, geometry, high performance building envelope, daylighting, solar control, shading, natural or mechanically assisted ventilation [27, 31, 34]. The effectiveness of passive strategies is maximized when they are implemented as part of an integrated, climate-responsive design process. Parametric design studies have demonstrated synergistic combinations of envelope performance, solar control, and ventilation strategies can achieve significantly greater energy reductions than isolated measures [34].

### 2.1.2 Active Design Strategies

After passive measures have reduced intrinsic energy demand, active systems are deployed to efficiently meet remaining loads while maintaining occupant comfort. In NZEBs, active design strategies focus on high-efficiency systems, climate-appropriate selection, and synergistic integration with passive design features. Common strategies include high efficiency HVAC systems, energy recovery ventilators (ERV) systems, energy efficient lighting, optimized equipment load sizing, as well as district scaling of active systems to improve energy efficiency [31, 35-38]

Table 1 illustrates that both passive and active designs play complementary yet distinct roles in enabling NZEBs. Passive designs are critical in delivering long term performance benefits independent of operational complexity, while reducing the scale and cost of downstream active systems. Active designs address the remaining energy demand. The effectiveness of active systems is contingent on prior passive demand reduction and are subsequently reliant on optimized operational strategies to function as intended.

### 2.2 Renewable Energy System

Real-world data is essential for research in BEMS because it offers accurate insights into actual building performance, supporting the development of effective and applicable energy management strategies. The examples listed below are real-world datasets in BEMS that have been curated to support research in energy monitoring, system optimization, and the development of intelligent control strategies for improving building performance and sustainability.

The notion of NZEB signifies a substantial transition towards environmentally friendly building methodologies, incorporating renewable energy (RE) systems such as photovoltaic (PV) and wind power, in conjunction with vital energy storage solutions (ESS). The structures are defined by their ability to produce an equivalent amount of energy from renewable sources as they utilize annually, thereby achieving energy stability and reducing dependence on external resources. The integration of onsite RE technologies, including PV panels and wind turbines, enhances building autonomy and advances overarching ecological goals in urban areas. The emphasis on self-sufficiency and diminished carbon emissions closely corresponds with current global initiatives to address climate change and improve energy security, as shown in Figure 3. The summary Table 2 shows contemporary literature on BEMS integration with renewable energy technologies, including onsite PV generating and battery storage, along with their evaluation metrics, gaps, and findings. This emphasis on self-sufficiency and carbon reduction aligns with global climate change and energy security efforts. For example, [39] examine improved semi-transparent PV glass, illustrating how Building-Integrated Photovoltaics (BIPV) offer visual and functional advantages while facilitating energy generation in buildings.

Table 1 Key passive and active design strategies supporting NZEBs performance.

| Design category | Strategy / technology | Key references | Key findings relevant to NZEBs |
|---|---|---|---|
| Passive | Building orientation and form | [30, 32] | Early-stage geometry and orientation significantly influence solar gains and thermal losses, with building-scale parameters showing stronger correlation with net energy performance than urban-scale form. |
| Passive | High performance building envelope (insulation, airtightness, glazing) | [27, 30, 33] | Envelope optimization is the dominant driver of heating and cooling load reduction across climates, enabling substantial demand reduction prior to system selection. |
| Passive | Daylighting, solar control and shading | [27, 34] | Integrated shading and solar control strategies reduce cooling loads while maintaining daylighting performance, particularly in cooling-dominated climates. |
| Passive | Natural and hybrid ventilation | [27, 31, 34] | Climate-responsive ventilation strategies can achieve energy savings ranging from approximately 13% to over 60%, but performance is constrained in hot–humid contexts without moisture control. |
| Passive | Integrated smart design | [30, 34] | Synergistic combinations of climate-aware passive measures outperform isolated strategies, emphasizing the importance of early-stage integrated design. |
| Active | High-efficiency HVAC systems | [35, 36] | HVAC systems remain the largest energy consumer in NZEBs; optimal system selection is climate-dependent and strongly influenced by prior load reduction. |
| Active | Energy Recovery Ventilation (ERV) | [31, 36] | Heat recovery ventilation reduces HVAC energy consumption by approximately 13–19%, with effectiveness varying by climate and humidity. |
| Active | Efficient lighting and appliances | [35, 36] | Reduction of internal loads lowers both electrical demand and secondary cooling loads, influencing HVAC sizing and net energy balance. |
| Active | System integration and optimal sizing | [30, 37] | Coordinated passive–active design prevents system oversizing and performance gaps, improving efficiency and cost-effectiveness. |
| Active | District scaling (district cooling/ heating) | [38] | District-scale cooling systems can enhance efficiency and peak load management in dense urban contexts, complementing building-level NZEB strategies. |

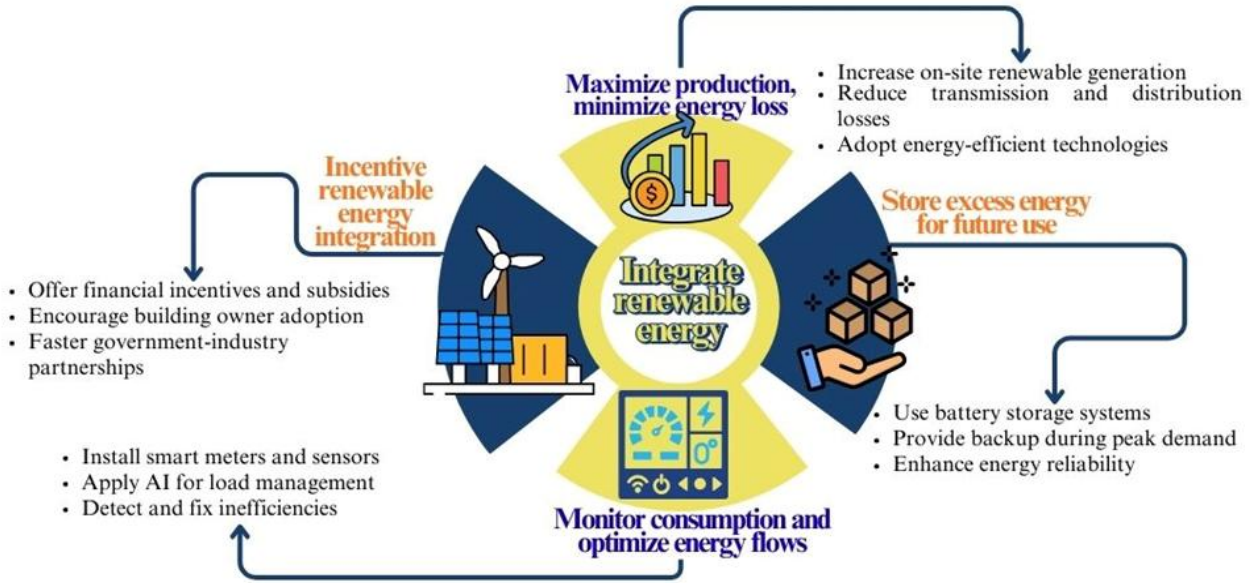

Fig. 3 Integration of renewable energy technologies for achieving net-zero energy buildings

A systematic approach to the integration of RE technologies is imperative to attain net-zero status. This entails the optimization of the configuration and operation of RE systems to maximize energy production and mitigate losses, in addition to the selection of suitable systems. For example, [41] reviewed real-time management of PV and battery microgrids, which enhances energy savings in apartment buildings, while [42] presented hybrid RE systems (solar, wind, and battery) controlled by LSTM-ANN models to achieve efficient smart-grid operation. Table 2 illustrates that a variety of studies have detailed strategies that capitalize on the on-site installation of renewable technologies and underscore the significance of ESS in regulating the intermittency of these resources. [43] demonstrates that intelligent automation in BEMS contributes to optimal PV-based energy management in residential and office buildings.

In addition, the integration of RE into NZEB is enhanced using BEMS. Real-time and data analytics can be employed by BEMS to monitor energy production and utilization, thereby enabling the dynamic adjustment of energy flows to optimize consumption patterns. Table 2 also delineates key outcomes from a variety of studies, demonstrating how BEMS strategies contribute to the overall efficiency and functionality of NZEBs by supporting the management of onsite generation and storage. [40] underlines the role of smart grids in reinforcing sustainable energy management, though they also identify persistent gaps in the integration of AI with renewable systems. The systems can intelligently manage loads, anticipate energy requirements, and ensure that RE is utilized efficiently by integrating AI-driven predictive and prescriptive analytics [43].

Eventually, the shift to NZEBs necessitates both technological advancements and supportive policies and incentives that promote investment and execution. The extensive summary study provided in Table 2 underscores existing deficiencies in comprehension and scalability, while pinpointing many building categories, from residential to commercial, that may benefit from these renewable systems. Governments, industry stakeholders, and researchers are required to collaborate to establish and enhance regulatory frameworks that facilitate the incorporation of RE solutions into the design of buildings. By establishing ambitious objectives and offering financial incentives for developments that incorporate onsite renewable technology and energy efficiency initiatives, stakeholders can promote the extensive use of NZEBs, facilitating a more sustainable future.

2.3 Operational Strategies

While passive and active design strategies, in combination with renewable energy supply, form the technical feasibility and energy balance of NZEBs, sustained net zero performance depends on how buildings are operated. Operational strategies address the temporal mismatch between energy demand, renewable generation, and grid conditions; ultimately determining whether net zero targets can be reliably achieved with consistency in real-world conditions [44-45].

Literature review highlighted energy flexibility as a central requirement for NZEBs. Energy flexibility refers to NZEB's ability to adapt its energy demand and on-site energy generation in response to external inputs such as renewable or energy storage availability, grid constraints, while maintaining indoor air quality (IAQ) and occupant comfort [44, 46].

Demand side management (DSM) strategies leverage on HVAC systems, thermal and energy storage, and controllable appliances to modulate demand profiles with variable renewable generation. Studies showed that load shifting and scheduling play a critical role in NZEB performance. Coordinated scheduling of HVAC operation, storage charging, and flexible loads can significantly improve renewable self-consumption and reduce reliance on grid imports, particularly in buildings with on-site PV and battery systems [47-48]. However, these benefits are constrained by comfort requirements, storage capacity, and user behavior, and simplistic rule-based approaches often fail to achieve robust performance under changing conditions [49].

Peak demand reduction and grid interaction have also emerged as important operational objectives. NZEBs increasingly interact with the grid not only as energy consumers but also as flexible resources capable of peak shaving, load modulation, and export management [44], [50]. Studies incorporating electric vehicles, storage, and multi-energy systems further demonstrate that grid-aware operational strategies can enhance both system-level efficiency and building-level performance [51].

BEMS serve as the primary coordination layer enabling these operational strategies. BEMS integrates monitoring, control, and scheduling functions across building systems and energy assets, translating high-level operational objectives into actionable control decisions [52]. Empirical studies indicate that well-implemented BEMS can deliver measurable improvements in energy performance and flexibility, while poor data quality, limited interoperability, and system complexity often undermine expected benefits [47-48].

Despite promising technologies, simulations and pilot studies, real world operational performance gaps remain. NZEBs frequently underperform due to occupant behavior, maintenance shortcomings, degradation of system performance, and the increasing complexity of flexible operation [53-54]. These findings highlight that operational strategies must be considered an integral component of NZEB design and delivery.

2.4 Building Certification, Codes and Standards

The implementation of NZEB is strongly influenced by the regulatory, institutional, and certification frameworks within which buildings are designed and operated. While design strategies, renewables and operational strategies address the technical and how-to's in enabling net zero performance; institutional certification systems, building codes, and standards provide the formal mechanisms through which NZEB concepts are defined, verified, and adopted in practice. These frameworks vary across regions in terms of scope and enforcement, contributing to uneven uptake of NZEBs globally.

At the international level, early NZEB concepts emerged primarily through research-driven initiatives and demonstration projects. [57] report documented a range of early net zero case studies. These early efforts preceded formalized certification or regulatory requirements and relied on project-specific definitions and performance metrics.

Over time, supranational and national directives such as the European Union's amended Energy Performance of

Table 2 Recent studies on renewable energy systems for NZEB

| BEMS Strategies | Approach (PV, wind) | Technology | Method | Evaluation metrics | Key outcomes | Gaps | Building types |
|---|---|---|---|---|---|---|---|
| Energy efficiency [55] | PV generation (review) | Solar PV | Design reviews | Energy consumption indices | Identifies approaches to energy reduction | Need for investment access and better financial models | Residential |
| Optimal energy management [41] | On-site PV generation | Solar PV, battery (Microgrid) | Real-time management | Energy savings | Enhances real-time energy management in apartment buildings | Limitations in scalability to multifamily units | Residential |
| Building-integrating microgrid (BIM) for NZEB [39] | On-site PV and Wind | Solar PV, Wind Turbines | Aperiodic micropatterns method | Enhanced Energy Efficiency | Lower energy demands | Limitations in integration | Residential, Commercial |
| Phase change materials (PCM) based thermal storage [56] | PV, wind generation (review) | PCM | Modelling & control analysis | Carbon neutrality metrics | Enhancing building energy efficiency for carbon neutrality | Required for field studies | Commercial |
| Intelligent automation for BEMS [43] | PV Onsite intelligent automation | Intelligent automation | Multi-criteria decision model | Sustainable energy management | Enhances decision-making for optimal energy management | Limited energy sources and technologies | Residential, offices |
| Energy management with hybrid RE sources [42] | On-site control system with PV, wind for Microgrid | Hybrid RE (solar, wind, FC, battery) | Design & control of LSTM-ANN | Efficiency performance | LSTM-ANN controllers for smart grid energy management | Require real-world validation | Smart buildings |
| Smart grid for sustainable energy management [40] | Smart grid | Smart grid | Systematic review | Sustainable energy management | Address integration issues of AI with renewables | Significant gaps in implementation work | Residential, Commercial |

Buildings Directive (EPBD) started to institutionalize the concept of nearly zero-energy buildings (nZEB) by requiring member states to adopt minimum energy performance standards and long-term renovation strategies Meanwhile, international research programs led by the International Energy Agency (IEA) have developed foundational definitions, performance boundaries, and evaluation methodologies for net zero and energy-flexible buildings, notably through IEA EBC Annex 40 and Annex 67 [44, 57].

More recent standards such as the UK Net Zero Carbon Buildings Standard, Canada's Zero Carbon Building (ZCB) Standard, and ASHRAE Standard 228 reflected a shift toward outcome-based metrics, emphasizing measured energy use, peak demand, and whole-life carbon performance [62-64].

Regionally adapted voluntary green building rating systems such as Singapore's Green Mark, Australia's Green Star, and Malaysia's Green Building Index (GBI) demonstrated localized application of net zero initiative contextualized to local climates, grid conditions, and market readiness [65-67].

Collectively, these certification systems and standards provide the institutional backbone for global NZEB development and deployment. However, variation in definitions, accounting methods, and verification rigor highlight the need for harmonization. Table 3 illustrate net zero related codes, standards, and certification frameworks in major developed economies, with Malaysia's GBI certification added for contextual local comparison.

**3.0 IoT enabled Integration of BEMS in NZEB**

IoT enabled integration extends the functional capability of BEMS in achieving NZEB performance by facilitating ubiquitous sensing, real-time connectivity and data informed decision making. Through the deployment of IoT applications such as sensors, controllers, actuators and smart meters, BEMS is able to improve energy efficiency by continuously monitoring energy consumption, indoor environmental conditions, and on-site renewable energy generation at high spatial and temporal resolution, thereby enhancing overall energy efficiency.

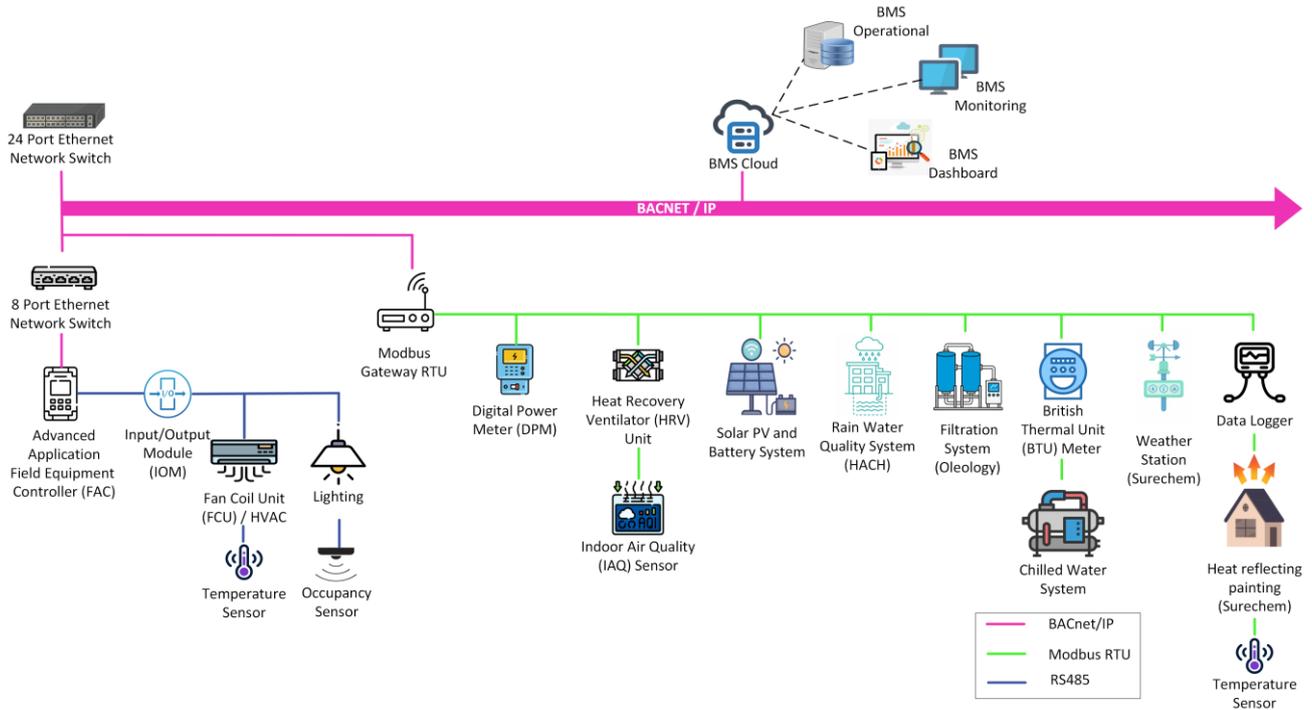

Figure 4: example of IoT application and network architecture in BEMS.

Effective communication architecture in BEMS is fundamental to the reliable system. Therefore, BEMS adopts layered network communication that integrates physical, field, network and application-level protocols to support interoperability and scalability of the BEMS system. Figure 4 illustrates example of IoT application and network architecture in BEMS.

3.1 IoT Application

Numerous research has been conducted on the implementation of BEMS technologies and tools by integrating IoT, AI, sensors, PV, and HVAC systems to enhance energy efficiency in NZEBs. The research highlights the role of BEMS in NZEB in achieving sustainable, cost-effective, and environmentally friendly optimization.

[68] has proposed PV technology integrates with an IoT based control mechanism to optimize energy generation and consumption. The methodology involves deploying sensors, motion detectors, and a Raspberry Pi-based gateway to monitor ambient light, user movements, and historical energy usage data, allowing for automated control of lighting and other energy consuming devices. The system also includes a web application that enables users to interact with energy settings remotely and a cloud-based database for real-time energy analysis. To improve energy efficiency, the system automatically adjusts lighting based on ambient conditions and exports excess solar-generated electricity to the grid. The results demonstrate that the IoT based BEMS significantly reduces energy consumption, enhances sustainability, and offers economic benefits, making it a viable approach for nearly zero-energy buildings [68].

[69] explores the application of HVAC systems in BEMS by analyzing real-time data from smart sensors and automated control systems to optimize energy efficiency and indoor climate conditions. By leveraging data-driven approaches, the research demonstrates how BEMS can enhance HVAC performance, detect inefficiencies, and reduce overall energy consumption in commercial buildings. [70] proposed smart HVAC system, equipped with fuzzy logic controllers, sensors, and automated air conditioning, operates through a ZigBee-based M2M network for real-time energy optimization. A real-world prototype tested in laboratories confirmed its effectiveness

Table 3 Major net zero related codes, standards, and certification frameworks

| Codes / Standard / Framework | Type | Scope & Geography | Primary Focus | Relevance to NZEB |
|---|---|---|---|---|
| IEA EBC Annex 40 [57] | Technical report / framework | International | Net zero energy definitions, solar integration, system boundaries | Widely referenced conceptual framework for NZEBs |
| IEA EBC Annex 67 [44] | Technical report / framework | International | Energy Flexible Buildings, demand response, grid interaction | Emphasize NZEB operation with energy flexibility |
| EU EPBD (Directive 2018/844) [58] | Mandatory regulation | European Union | Mandatory nZEB requirements, long term renovation strategies | Policy driver for large-scale NZEB adoption in EU member states through national law. |
| LEED Zero [59] | Voluntary Certification | United States - International | Verified net zero energy, carbon, water, waste | Performance based LEED add on for NZEB verification |
| BREEAM (BRE (UK)) [60] | Voluntary Certification / Rating | United Kingdom - International | Comprehensive, policy and planning aligned | A pioneering green building sustainability rating system developed by BRE (UK) frequently referenced by local authorities, public sector and commercial developments. |
| DGNB Carbon Neutral Framework [61] | Voluntary Certification | Germany - European Union | Whole life carbon neutrality, lifecycle assessment | An extension of DGNB sustainability certification system developed by German Sustainable Building Council with emphasis on carbon accounting |
| UK Net Zero Carbon Buildings Standard (Pilot rev 2) [64] | Voluntary National Standard | United Kingdom | Unified national definition, operational & embodied carbon | Sets clear requirements and reporting guidelines for buildings to be classified as Net Zero Carbon Aligned |
| ASHRAE Standard 228 [62] | Voluntary Technical Standard | United States - International | Performance based building energy modeling & verification | Engineering focused, performance and metrics driven enabler for consistent NZEB performance assessment |
| Canada Zero Carbon Building (ZCB) Standard v3 [63] | Voluntary National Standard | Canada | Operational & embodied carbon neutrality, grid interaction | Comprehensive carbon-based frameworks developed by CAGBC are increasingly required for federal projects. |
| Singapore Green Mark [65] | Quasi Mandatory National Standard / Rating | Singapore | Tropical high-performance buildings, energy & carbon | A national green building scheme embedded in regulation, often required via Building Control Act |
| Australia Green Star Standard [66] | Voluntary Certification / Rating | Australia | Net zero carbon, health, resilience | Australia's national voluntary green building rating system, developed by the Green Building Council of Australia (GBCA) |
| Malaysia Green Building Index [67] | Voluntary Certification / Rating | Malaysia | Energy efficiency, tropical design, renewables | Malaysia's original and well-established green rating tool. |

in reducing energy consumption, stabilizing indoor temperatures, and enhancing building sustainability.

[71] highlights the role of smart sensors, wireless sensor networks (WSNs), cloud technology, and IoT integration in monitoring and optimizing energy consumption in buildings. BEMS components in the systems include HVAC systems for climate control, PV panels for renewable energy generation, IoT based lighting and occupancy sensors, and automated energy management platforms to reduce waste and improve efficiency. This system proposes a smart building template, integrating real-time data collection, automated controls, and predictive analytics to enhance building sustainability and performance.

Smart meters are essential components of BEMS enabling real-time energy monitoring, self-consumption analysis, and bidirectional energy trading. In addition, smart meter is able to remotely on/off the appliances to avoid energy wastage. In NZEBs, smart meters help evaluate self-consumption efficiency by analyzing real-time electricity load consumption and PV generation data [72]. The study highlights that precise time resolution in smart meter data collection is crucial to accurately matching PV generation with building energy demand, avoiding overestimated self-sufficiency indices. Overall, smart meters enhance energy efficiency, sustainability, and cost optimization in BEMS by improving data accuracy, demand-side management, and renewable energy integration.

[73] proposed a comprehensive data collection methodology for BEMS by integrating real-time electricity consumption monitoring, environmental sensing, and user behavior analysis. Smart meters are installed to track electricity usage at hourly, daily, and monthly intervals, capturing peak and valley power demands, while temperature and humidity sensors collect indoor environmental data to assess the correlation between energy consumption and environmental conditions. Additionally, the operational status of major energy-consuming equipment (e.g., air conditioning, lighting, elevators) is recorded, and residents' energy use preferences are gathered through questionnaires and smart home systems, allowing for a deeper understanding of human factors influencing energy consumption.

The implementation of HVAC systems in BEMS allows for automated climate control by integrating smart sensors, IoT, and AI-driven optimization, ensuring efficient heating, cooling, and ventilation based on real-time conditions. This integration is crucial in reducing energy waste and operational costs, as BEMS can predict and adjust HVAC performance according to occupancy patterns and external weather data. The importance of HVAC in BEMS lies in its ability to enhance occupant comfort, improve air quality, and contribute to sustainability efforts, making buildings more energy-efficient and environmentally friendly. Table 4 shows the implementation of BEMS components.

The implementation of BEMS varies across different global regions due to differences in objectives, approaches, and constraints. These differences are largely shaped by climatic diversity, as varying weather conditions significantly affect energy demand and building performance [74]. For example, in Asia, East Asia (e.g.: Japan, Northern China, South Korea) has a temperate four-season climate; South Asia (e.g., India, Sri Lanka, Bangladesh) experiences tropical conditions; Central Asia (e.g.: Iran, Kazakhstan, Uzbekistan) is characterized by a semi-arid climate; Southeast Asia (e.g.: Malaysia, Indonesia, Singapore) falls within the equatorial zone; and parts of Western Asia (e.g.: Turkey) have a Mediterranean climate [1, 75]. Likewise, European countries typically experience temperate, oceanic, and continental climates, whereas the Middle East and North Africa are predominantly arid. These climatic variations require tailored BEMS strategies that consider regional energy use patterns and environmental conditions.

In regions with arid and semi-arid climates, there is heavy reliance on air conditioning during hot, sunny periods, while temperate regions require cooling in summer and heating in winter [1]. Areas with prolonged summer seasons such as much of Asia and the Middle East can benefit from incorporating PV systems alongside HVAC systems in BEMS to reduce reliance on the power grid and enhance energy efficiency. To maximize energy performance and support the development of NZEB, BEMS should include advanced monitoring technologies, such as IoT systems, smart sensors, and real-time data analytics, for responsive and efficient energy control [76]. Ultimately, the success of BEMS in any region depends on its ability to adapt to the local climate and integrate appropriate technologies accordingly. Table 4 shows the implementation of BEMS across different global regions based on objectives, approaches, and constraints.

3.2 Interoperability and Smart Grid Integration

Interoperability of IoT applications in BEMS is very crucial in maintaining building effectiveness and system operability. It involves integration and connectivity of various IoT devices, network protocol and system within a building that allows real-time monitoring, control and data transmission to maximize energy efficiency and reduce energy wastage.

Several issues have been identified as a significant barrier to communication which contributes to a lack of interoperability of the systems. The major issues are on the heterogenous nature, for instance discrepancies in data formats and architecture have affected data exchange and integration process, leading to inefficiency [77]. This was supported by the lack of standard format in integrated IoT BEMS solutions that has made interoperability more challenging [78].

In other cases, interoperability has been hampered by different devices and systems from different vendors, usually with proprietary communication protocols, for example, although BACnet offers interoperability, it also has compatibility issues with modern IoT devices [19]. [79] has mentioned that unifying descriptive models that encompasses both functional roles and system features offered as a solution that can be applied across diverse devices and data types.

The selection and standardization of communication protocols are important to ensure interoperability across IoT-enabled BEMS. [80] recommended open standard and interoperability frameworks such as BACnet, MGTT and OPC UA are suitable to enable seamless communication and system integration BEMS. In addition, preserving with specific protocol in BEMS system may help increase operability, for example Ethernet, WiFi and RS485 may apply at physical data transmission, while BACnet and Modbus are specified for device level integration and control at filed level protocols [18]. [81] also highlighted scalable and interoperable architecture should be prioritized in future research to overcome constraints in data communication between different IoT devices and platforms which has led to fragmented systems.

Aligned with the energy management requirement, the integration of BEMS with smart grid (SG) is essential for overall system and faces its own interoperability difficulties. Integration of SG in BEMS system implement demand side management practice where it is allowing interaction between building administrator, customer and utilities for monitoring and regulation in building energy utilization [82]. This was supported by [83] the importance of reinforcing end-customer participation and optimizing grid potential to ensure the effectiveness of the program and enhance grid flexibility and reliability.

The diversity of communication protocols in BEMS has made interoperability an arduous task. However, the application of standardized protocols such as IEC 61850, IEC 61970/61968 Common Information Models, IEEE 1815, and IEEE 2030.5 are essentials to provide smooth communication and data management within SGs [84]. These standards provide framework for consistent data exchange, system integration and coordinated control within heterogeneous SG infrastructure.

3.3 BEMS Communication Standards and Protocols

Open protocol serves a key function in BEMS to ensure seamless communication and interoperability between different components are achieved. The purpose of communication standards and protocol in BEMS are to ensure interoperability, systemized monitoring and control, efficient data exchange, reducing deployment efforts and costing and enable advanced analytics. Variety of BEMS communication protocols based on the application evolving from older, established, to lightweight IoT protocols.

There are several established communication protocols identified due to reliability, robustness and widespread adoption to automation building. BACnet is designated for building automation where it provides a standardized communication protocol that enables interoperability across various building automation subsystems including HVAC, lighting, sensors, security and fire safety and access control [85-86]. It operates over existing physical and network standards, for example RS485, Ethernet and MS/TP (Master Slave/Token Passing) [87], while BACnet introduces security features like encrypted communication and device authentication. Other than that, Modbus is also commonly used in industrial control systems and building automation and implemented Modbus TCP (over ethernet) or Modbus RTU (over serial line), but Modbus TCP experienced lack proper authentication mechanism that may risk cyberattacks [88]. LonWorks also offered open standards protocol that used both data protocol and electrical standards based on the LON schemes [89]. Although LonWOrks is no longer a popular choice, it still exists in many operational buildings. KNX or International Standards ISO/IEC 14543-3) is also widely used in Europe specifically in commercial and residential building [90].

While the conventional protocol is still in use, lightweight IoT protocols are gradually taking place to provide efficient data transfer in modern connected with

BEMS. Lightweight IoT protocols was introduced as low power consumptions, scalability, and reliable in unstable networks [91]. MQTT has offered efficient data transfer across limited bandwidth networks [92], as well as CoAP. However, a Low Power Wide Area Network (LoRaWAN) technology proposed connectivity in large scale smart buildings with low power and long-range communication for energy management application [93].

Other than that, there are many other communications protocols integrated with BEMS application for example Ethernet that supported high-speed data transmission and frequently carried protocols like BACnet/IP [18] and Dali applied in lighting [94]. DeviceNet, C-bus, m-bus also widely adopted in building automation [95] along with EnOcean that is a standard based on IEEE 802.15.4 [96].

To ensure effective BEMS communication, open standards protocol and semantic models should be prioritized to sustain interoperability across diverse devices while reducing integration requirements. Furthermore, the adoption of secure protocol with built-in encryption and authentication together with robust security measures to protect against cyber threats.

3.4 Overview of Real-Worls Datasets in BEMS

Real-world data is essential for research in BEMS because it offers accurate insights into actual building performance, supporting the development of effective and applicable energy management strategies. The examples listed below are real-world datasets in BEMS that have been curated to support research in energy monitoring, system optimization, and the development of intelligent control strategies for improving building performance and sustainability.

[97] presented comprises six years (2018–2024) of continuous measurements collected from electricity, heating, and cooling meters, as well as a weather station installed at the Honda R&D Europe facility located in Offenbach am Main, Germany. Data acquisition was conducted using a variety of specialized metering devices and sensors, with temporal alignment accounting for local timezone variations, including adjustments for daylight saving time. Throughout the data collection period, various disruptions including measurement outages, maintenance activities, and device replacements that necessitated a comprehensive data cleaning and post-processing protocol. A structured seven-step pipeline was implemented to identify and correct anomalies, harmonize data formats and naming conventions, perform time alignment, resample the data into uniform intervals (1 minute, 15 minutes, and 1 hour), and generate derived measurements where necessary. Additionally, a reduced version of the dataset is made available, offering an aggregated and less complex representation of building energy consumption, energy production (electricity, heating, and cooling), and associated environmental conditions, thereby facilitating broader applicability in research contexts.

[98] introduced an energy consumption monitoring dataset from the Hong Kong University of Science and Technology (HKUST), comprising data from over 1,400 meters across more than 20 buildings, collected over a two-and-a-half-year period. The dataset was curated using the Brick Schema, ensuring semantic consistency and transforming raw measurements into a research-ready format. It enables a wide range of applications, including load pattern analysis, fault detection, demand response planning, and energy consumption forecasting. This dataset was collected to address the growing need for accurate electricity management in campus environments, where understanding load patterns is essential for enhancing energy efficiency and optimizing usage. However, the availability of detailed electricity load data for campus buildings and their internal systems remains limited, hindering progress in related research.

[166] collected datasets for over three years datasets from an office building in Berkeley, California, using over 300 sensors to record energy use, HVAC performance, environmental conditions, and occupancy across two office floors. A three-step curation process was used to clean the raw data, model system metadata with the Brick schema, and describe metadata using a semantic JSON schema. The resulting research-grade dataset supports applications such as energy benchmarking, load analysis, predictive modeling, and HVAC optimization to enhance building efficiency and reduce energy use and emissions.

The availability of high-quality real-world data is critical to the advancement of BEMS, enabling precise system modeling, performance analysis, and the design of intelligent control strategies. While recent large-scale datasets have supported diverse applications, they often present challenges such as missing data, sensor malfunctions, and high acquisition costs. To address these issues, synthetic data that is generated based on validated real-world datasets offers a valuable complement, enhancing research flexibility and supporting machine learning development. The integration of both real and synthetic data will be essential for building scalable,

Table 4 BEMS implementation across different global regions based on objectives, approaches, and constraints.

| Components/ Application | Objectives | Approach | Measured Parameter | Limitations | Region/ Climate | Reference |
|---|---|---|---|---|---|---|
| PV | Enhance energy efficiency by incorporating solar PV | Uses IoT and automation to balance energy generation and consumption in smart buildings. | Power consumption, (Watt, W), Electricity consumption (kW), Monthly energy consumption (kW), $CO_2$ emission (kg) | Weather-dependent energy generation requires effective energy storage solutions. | Tehran, Iran / Asia / Semi-Arid | [68] |
| IoT based control system | Optimize energy consumption through real-time monitoring and control. | Uses a set of sensors, servers, and wireless networks to track and manage energy usage. | | Requires stable network infrastructure; security and data privacy concerns. | | |
| Wireless Sensor Network | Collect and transmit environmental and energy-related data in buildings. | Deploys a network of interconnected sensors (e.g., temperature, motion, energy meters) to gather data. | | High energy consumption of sensor nodes; potential data loss due to interference. | | |
| HVAC | Optimize HVAC for energy efficiency. | Uses IoT-enabled sensors and AI-based optimization models to adjust HVAC settings dynamically. | | High computational requirements; latency issues in real-time response. | | |
| Smart meter (SM), sensor | Monitor and collect real-time energy usage data | Use IoT-enabled sensors to track temperature, occupancy, and energy consumption | Monthly energy consumption (kWh), Peak energy consumption before and after optimization (kWh), Energy efficiency (%) | High installation cost; Data security concerns | China / Asia / Temperate (4 seasons) | [73] |
| AI-based Optimization | Improve energy efficiency by predictive control | Machine learning models analyze patterns and optimize HVAC, lighting, and appliances | | Requires extensive training data; Complexity in implementation | | |
| Automated HVAC control | Optimize HVAC efficiency | AI-based predictive control and scheduling | Energy consumption (KWh), Outdoor/indoor temperature (C°), Operational HVAC (°F) | Complexity in real-time adaptation, initial setup cost | Barkeley, California / Western US / Mediterranean | [69] |
| SM & sensors | Measure and monitor real-time energy consumption | IoT-based sensors and SM collect data for analysis | | High cost, data privacy concerns, and integration challenges | | |
| Predictive Maintenance | Prevent failures and improve system longevity | AI-driven fault detection and predictive analytics | | High data processing requirements, sensor reliability issues | | |
| Smart sensors and actuators | To monitor and control energy consumption. | Automated energy monitoring through IoT-enabled smart sensors. | Ambient temperature (°C), Energy consumption with/without M2M (kWh) | High initial cost | Sakarya, Turkey / Asia / Mediterranean | [70] |
| Cloud-based data processing | For real-time energy management. | Data analytics and cloud computing for predictive energy management. | | Data security concerns, since cloud-based energy management may be vulnerable to cyber threats. | | |
| Communication networks (M2M technology) | To enable seamless data exchange between devices. | Wireless communication protocols (e.g., ZigBee, Wi-Fi) to enhance system connectivity. | | Network dependency, as any failure in the communication system could affect energy management. | | |
| Cloud technology | To improve energy efficiency in smart buildings through the application of IoT technologies | To collect data from various building systems and store this information in a cloud database. | Energy consumption (Load, PV) (kWh), PV sufficient (kWh), Percentage error (%) | May not be feasible or legally applicable in real-world smart cities | Greece / Europe / Temperate (4 seasons) | [71] |
| Wireless Sensor Networks | Monitoring and optimizing energy consumption in buildings | Management system uses this data to track energy use, detect energy waste, and evaluate how the building performs on smart readiness indicators | | | | |
| PV | To create a friendly and energy-efficient indoor environment by adjusting conditions based on occupants' needs. | Utilize a clean and renewable energy source, helps to reduce greenhouse gas emission | | | | |
| SM, PV | Use smart meters to facilitate a transition to a low-carbon economy by enabling PV self-consumption, especially in NZB | Using SM to record load consumption and PV generation profiles at different recording intervals and reporting periods. | | The study considers PV systems from 0.01 to 10 kWp, which reflects typical household sizes. Household energy use is unpredictable, as it depends on occupant behavior and appliance usage | Spain / Europe / Mediterranean | [72] |

Table 5 The types of datasets generated for BEMS.

| Category | Parameter | Details / Units | Ref. |
|---|---|---|---|
| Energy Consumption Data | Interval-based measurements | 10 minutes, hourly, daily | [98-100] |
| | Load types | HVAC (heating/cooling), Lighting, Plug loads, Elevators, Mechanical systems | [69, 100, 101] |
| | Spatial breakdown | Zone-level or floor-level (office, lobby, meeting room) | [98, 102] |
| Indoor Environmental Data | Temperature | °C or °F, by zone | |
| | Humidity | % Relative Humidity (%RH) | [69] |
| | $CO_2$ concentration | Parts per million (ppm) | |
| | Light levels | Lux | |
| | Noise levels (if relevant) | Decibels (dB), if applicable | |
| Occupancy Data | People count | Number of people per room or zone | [69] |
| | Motion sensors | Binary (0/1) triggers | |
| | Occupancy schedules | Example: 9am–5pm weekdays | |
| External Weather Data | Outdoor temperature | °C or °F | [103] |
| | Humidity | %RH | |
| | Solar irradiance | W/m² | |
| | Wind | Speed (m/s), Direction (degrees or cardinal) | |
| | Cloud cover / weather condition | Coded values or descriptions | |
| Equipment / System Control States | HVAC status | ON/OFF or % load | |
| | Fan speeds / valve positions | % open or control values | [100, 104] |
| | Lighting control levels | Dimming percentage (%) | |
| | Energy storage / PV generation states | ON/OFF, % charge, generation output | |
| System Metadata | Zone type | Open office, Laboratories, Hallway, Big Classroom, Small Classroom etc. | [98, 102] |
| | Floor area | m² | |
| | Construction type | Lightweight, heavy, insulated | |
| | System capacity | e.g., chiller kW, lighting power density (W/m²) | |

robust, and future-ready energy management solutions. Notably, the dataset by [98] aligns with the scope of our project, as it also focuses on analyzing energy consumption in academic buildings comprising classrooms, laboratories, and offices.

### 3.5 BEMS Dataset Generation

BEMS acquires data from sensors on energy consumption, equipment operation schedules, occupancy patterns, and indoor environmental parameters such as temperature, humidity, and air quality, which are then utilized for optimization, predictive control, and informed decision-making to enhance energy efficiency and occupant comfort. The types of datasets generated for BEMS are as in Table 5.

### 3.6 AI Implementation in BEMS and Related Works

Recent advancements in BEMS highlight the integration of cutting-edge AI, ML, and deep reinforcement learning (DRL) techniques to optimize energy usage, enhance system efficiency, and reduce environmental impacts.

The DRL-based BEMS proposed by [101] focuses on optimizing energy consumption, heat management, and carbon emissions in residential buildings by integrating power-to-heat (P2H) technology and a two-stage heat pump system. It employs a novel dynamic action subset–twin delayed deep deterministic policy gradient (DAS-TD3) algorithm, which enhances decision-making under uncertainty by efficiently controlling distributed energy systems. The system collects and monitors data such as electricity usage, thermal demand, carbon emissions, weather conditions, PV generation, and user preferences to inform the DRL agent's actions. Results show that the proposed system outperforms conventional approaches, achieving reduced energy consumption, lower emissions, and improved thermal comfort [101].

Complementing this approach, [105] explore Artificial Adaptive (AA) systems, specifically using LSTM networks, to predict energy consumption in smart buildings. Their method integrates statistical tools such as PCA, ARIMA, and autocorrelation analysis to preprocess data and capture temporal patterns. The Long-Short Term Memory (LSTM) models trained on historical energy usage, weather, and occupancy data achieved high prediction accuracy, with a peak precision of 74%, RMSE of 0.08, and MAPE of 0.13. This study underlines the promise and complexity of deploying AA systems for real-time energy forecasting and control [105].

[103] proposed the use of Tabular GANs to generate synthetic electricity consumption data that mirrors real-world patterns, supporting smart city planning and energy management amid growing automation and Net-Zero goals. Using Python and the 'tabgan' library, the authors trained their model on real datasets to replicate the statistical features of actual electricity usage. The model utilized the 'Low Carbon London' dataset and NASA weather data to build realistic consumption profiles.

Results demonstrated that the synthetic data closely resembled the original dataset in statistical behavior and consumption trends. The study concludes that Tabular GANs are a promising tool for simulating electricity demand where real-world data may be limited or sensitive.

[98] provide a large-scale, high-resolution dataset to expand the accessibility of real data from the Hong Kong University of Science and Technology, capturing over 2.5 years of energy consumption across more than 20 buildings. Structured using the Brick Schema, this dataset facilitates integration with BEMS for applications such as load pattern analysis, fault detection, and demand response, supporting institutional energy performance evaluations.

[99] introduce BiTSA, a system designed to help build managers optimize energy consumption and achieve sustainability goals through an interactive, user-friendly dashboard. BiTSA integrates with Building Management Systems and supports advanced forecasting models such as DLinear, PatchTST, Informer, iTransformer, and GPT-2-based One-Fits-All, trained using the Adam optimizer. It processes historical building data which resampled to 10-minute intervals and preprocessed with second-order polynomial interpolation from the BTS-B and BLDG datasets. The system offers dynamic visualizations that provide actionable insights, enabling users to make informed and timely energy management decisions.

[102] propose a conditional diffusion model designed to generate high-quality synthetic energy consumption data by leveraging metadata such as building types, meter types, and geographic location. The model utilizes U-Net architecture with a time-embedding layer to effectively capture temporal dependencies and reverse a noise injection process to produce realistic data samples. It is trained on a dataset of 1,828 power meters from various global locations and outperforms traditional models like Conditional GANs and VAEs in statistical similarity and diversity. The results demonstrate a 36% improvement in Fréchet Inception Distance and a 13% decrease in Kullback-Leibler divergence, highlighting the model's capability in generating variable and realistic synthetic energy patterns.

For residential energy systems, [106] examine the feasibility of integrating Distributed Generation (DG) into a smart home environment using an AI-powered IoT-based Home Energy Management System (HEMS). The system aims to enhance energy management by reducing the Peak-to-Average Ratio, lowering electricity bills, improving energy savings, and increasing consumer comfort. It likely employs J8 machine learning algorithms and the Weka API to analyze load behavior and generate intelligent energy-saving recommendations, supported by IoT communication protocols such as MQTT for scalability. The system monitors live load patterns, appliance energy use, DG output, grid status, and user preferences to demonstrate the benefits of AI-based DG integration.

[69] presents a three-year dataset that supports research on HVAC systems and building energy management by analyzing energy consumption, environmental conditions, and occupant behavior. The HVAC system in the office building studied includes underfloor air distribution (UFAD), rooftop units (RTUs), variable-speed fans, and hydronic heating coils, all managed by an Automated Logic (ALC) WebCTRL Building Management System (BMS). The dataset methodology involves collecting real-time data from 300+ sensors, monitoring HVAC operations, electricity usage, occupancy levels, and environmental factors like temperature, humidity, and $CO_2$ concentration. The dataset was curated using a three-step process: cleaning raw data, structuring it using the Brick schema, and organizing metadata in a semantic JSON format for enhanced usability. This dataset can be applied for building energy benchmarking, load shape analysis, HVAC fault detection, and predictive analytics, providing valuable insights for optimizing building energy efficiency.

[104] presents a case study from the Solar Decathlon Middle East (SDME), where AI-driven monitoring, PV integration, and automated load control significantly improved energy efficiency. The proposed system utilizes smart meters, power tags, and pattern recognition software to monitor real-time electricity consumption and optimize energy usage through AI-based automation. IoT-enabled smart energy management systems improve building performance by dynamically adjusting HVAC settings, lighting, and appliance loads based on occupancy patterns and energy demand.

[100] has presented details of electricity consumption and indoor environmental data from a seven-story office building in Bangkok, collected over 18 months at one-minute intervals. The dataset includes power usage of individual air conditioning units, lighting, and plug loads across 33 zones, as well as temperature, humidity, and ambient light levels recorded by multi-sensors. These findings enable zone-level, floor-level, and building-level load forecasting, HVAC optimization, and anomaly detection for energy efficiency improvements. The study

highlights the importance of smart metering and real-time monitoring in BEMS, demonstrating how granular data supports demand response strategies and reinforcement learning algorithms for energy control. Ultimately, the dataset provides valuable insights for building simulation models, energy-saving strategies, and predictive analytics, contributing to sustainable and efficient building operations.

These studies collectively demonstrate the transformative impact of AI [69, 99, 104, 106, 107], DRL [101], and synthetic data generation [102, 103, 105] on modern BEMS. By leveraging both real-time and synthetic datasets, integrating advanced forecasting algorithms, and enabling autonomous decision-making, researchers and practitioners are advancing toward more resilient, efficient, and sustainable building energy systems. The synergy of data-driven intelligence and smart system integration is key to achieving net-zero goals and responding effectively to the dynamic demands of energy management in the built environment. Table 6 shows AI implementations in BEMS and related works.

## 4 Case Study

The integration of BEMS into NZEB offers significant advantages by providing a simple, user-friendly platform capable of accommodating complex operational needs and unique specifications. BEMS enables monitoring and control of equipment in real time by the facilities services and maintenance personnel, ensuring efficient operation and timely interventions. In addition, it supports performance benchmarking and asset tracking to optimize long-term operational efficiency. The system's flexibility also allows users to access data from multiple devices, including PCs, tablet or mobile devices, thereby enhancing usability and decision-making in achieving net zero performance goals.

4.1 Global Examples of Successful NZEB with advanced BEMS

Six case studies involving usage of BEMS for NZEB globally. Several implementations further demonstrate the effectiveness of BEMS and IoT in achieving net-zero in commercial and university buildings.

4.1.1 Commercial and office buildings: The Wings, Brussels, Belgium, Vorum, Germany and ArchiCe, Hanza Tower, Poland

The Wings, a 50,000 m2 a mixed-use development in Brussels, Belgium comprising a hotel, gym, restaurants and offices have adopted the Metasys Building Automation System from Johnson Controls Inc. to optimize its daily operations. The integration of the BEMS has enabled instant access to real-time building performance data, while simultaneously reducing carbon emissions and lowering operating costs.

Implementation of BEMS system has enabled automated control and real-time monitoring of the HVAC system. Integrating sensors such as temperature sensors, occupancy sensors, indoor air quality sensors with scheduling assistants, optimizes the BMS data. With sufficient operation data collected from BMS system, the indoor relative humidity of The Wing's was maintained to approximately 50%, which abides the Environmental Protection Agency (EPA) requirement of 30-50% relative humidity in buildings [108]. Indoor relative humidity is important to prevent mold growth, which will affect the health and well-being of the residents, such as dry and itchy skin symptoms, tiredness, fatigue and drowsiness [109]. To ensure only authorized users/devices connect to the BEMS platform, network access control is implemented using IEEE 802.1X authentication protocol. Other than securing network access control, communication between the field devices, controllers and BEMS server occurs over a trusted and verified network.

Similar to the The Wings, a multi-purpose facility in Germany, Vorum uses BMS to provide comfort to its occupants. Vorum is a healthcare facility comprises of multiple buildings and has to accommodate many occupants. The BEMS system of Vorum, provided by ABB's KNX ClimaECO is designed to control multiple buildings, including treatment and consultation rooms, swimming pool, fitness area and administrative buildings. In Vorum, temperature control of any room does not limit the control of HVAC system but also includes control of the room blinds, to enhance operational efficiency.

ArchiCe, an office building located in Hanza Tower, Poland has implemented ABB's i-bus® KNX as the building and energy management system to reduce the energy usage while maintaining the comfort of their employees. The temperature of rooms was controlled by integration of digital sensors touch screen and switches throughout the office. Temperature control in the workspaces was conducted by monitoring the real time temperature, along with ventilation rate and automated control of the curtains.

Table 6 AI implementations in BEMS and related works.

| Study | Function | Methods | Data Collection & Monitoring | Results & Findings |
|---|---|---|---|---|
| [101] | Optimize energy, heat, and carbon in residential buildings via DRL-based BEMS with P2H integration | Uses DAS-TD3 algorithm for control under uncertainty | Monitors electricity use, thermal demand, carbon emissions, weather, PV generation, user preferences | Outperforms conventional systems, improving energy use, emissions, and comfort |
| [105] | Predict and optimize energy use with Artificial Adaptive Systems in smart buildings | Uses LSTM with PCA, ARIMA, autocorrelation; Weka API for ML | Data from sensors, actuators, energy use, occupancy, weather, PCA, SkySpark | Achieved 74% precision, RMSE 0.08, MAPE 0.13, $R^2$ = 65% |
| [103] | Generate synthetic electricity data for smart planning and energy management | Python + 'tabgan'; trained on real data to mimic patterns | Used Low Carbon London data and NASA weather variables | Synthetic data matched original well; supports use in limited/sensitive data cases |
| [98] | Provide standardized energy data for research and analytics | Used Brick Schema for semantic curation of raw data | Data from 1400+ meters across 20+ buildings, 30-min resolution | Enables load pattern recognition, fault detection, demand response; used in campus energy reports |
| [99] | Enable building managers to optimize energy and sustainability via dashboard | Integrated forecasting models (DLinear, Informer, GPT-2); Adam optimizer | Data from BTS-B & BLDG datasets, resampled at 10-min intervals | UI provides actionable visual insights, real-time interactivity |
| [102] | Generate diverse, high-quality synthetic energy data using metadata | Conditional diffusion model with U-Net and time-embedding | 1,828 power meters; includes location, meter/building type, weather | 36% FID and 13% KL improvement over GANs/VAEs; accurate and variable patterns |
| [106] | AI-based integration of DG in smart homes for better energy management | Likely uses J8 ML, Weka API, MQTT protocol | Tracks load behavior, appliance usage, DG output, grid state, preferences | Demonstrates AI-HEMS potential in reducing costs and increasing efficiency |
| [69] | Provide HVAC and building energy dataset for research and analytics | Brick schema + semantic JSON format for usability | 3 years of data from 300+ sensors: HVAC, occupancy, temp, $CO_2$, electricity | Enables benchmarking, fault detection, predictive analytics |
| [104] | Improve energy efficiency through AI monitoring and load control in SDME case | Uses pattern recognition, smart meters, power tags | Monitors real-time electricity, adjusts loads based on patterns | Shows AI and IoT boosts efficiency via automation in PV-integrated buildings |
| [100] | Enable smart BEMS via granular consumption and environmental data | Smart metering with multi-sensors; real-time monitoring | 18 months of 1-min data across 33 zones: HVAC, lighting, temp, humidity | Supports load forecasting, anomaly detection, RL for control |

4.1.2 Airports: Minneapolis-St. Paul (MSP) International Airport, Minneapolis-Saint Paul, United States

Airports are energy-intensive facilities that accommodate staff and passengers around the clock. MSP Airport is a public international airport, accommodating 37.2 million passengers per year. MSP Airport covers an area of 3 million square feet and presents challenges in maintaining a conducive environment.

To maintain comfort, safety and efficiency of an airport, BMS acts as the digital backbone to coordinate various control systems to operate the airport. MSP airports integrate HVAC control, lighting, plumbing and security using BEMS provided by Honeywell's Niagara Framework™ to enhance occupants' comfort and operational efficiency. In MSP, occupancy sensors were integrated with lighting and ventilation control to enable intelligent and energy-efficient operation of the building system. Presence or absence of people in the designated space were the factor in adjusting the lighting levels and ventilation rates. In a fully occupied room, lights will be activated, and air circulation is increased to maintain a comfortable environment. Once the room is unoccupied, the lighting will be dimmed or switched off, and the ventilation rate will be reduced to a minimal level. These steps are important in lowering the energy consumption in MSP airport. Occupancy-based control improves user comfort and reduces unnecessary equipment runtime, leading to longer equipment lifespan and lower maintenance costs.

As MSP airport is a large facility with multiple tenants, the metering for tenant billing has been implemented. Energy usage of each tenant can be monitored using BMS, allowing the operator to track energy usage patterns, detecting abnormal energy trends that may indicate equipment malfunction and identify any inefficiency. The practice promotes transparency and energy accountability between the MSP operators and tenants.

MSP airport also aims to reduce 25% of construction costs by transitioning from proprietary systems to open-system architecture. By adopting open protocols and interoperable system, equipment can be sourced from multiple suppliers, encouraging competitive pricing and eliminating restriction of a single manufacturer's technology. Cost efficiency, system scalability, and long-term sustainability can be enhanced.

4.1.3 Academic buildings: The University of British Columbia, Vancouver, Canada and The University of Nottingham, United Kingdom

BEMS has been used in operation of academic buildings in the University of British Columbia, Canada and the University of Nottingham, United Kingdom. Universities are often built in a large compound, comprised of multiple buildings and operate like a mini city.

BEMS integrates class scheduling and occupancy sensors to control heating and cooling of any designated rooms. Rooms will be cooled or heated when there is a class scheduled but the HVAC system will be switched off if there is no motion detected by the occupancy sensors. Even though there are multiple buildings to be controlled, BMS is able to monitor all the rooms in a single platform, thus providing an overall status at real-time. To estimate building occupancy and to control building's HVAC system, the location information from Wi-Fi connected devices are used.

The University of British Columbia, Canada uses a BMS with multiple solution providers which are Siemens, Delta/ESC and Johnson Controls Inc. There are 4,700 units of Siemens controllers, 1,430 units Delta/ESC controllers and 1,260 units Johnson controllers were deployed throughout the campus and monitored using a single platform. This open-protocol system allows controllers and sensors from different operators to interact with each other and send feedback to a single platform.

In the University of Nottingham, United Kingdom, operation system from Schneider Electric's EcoStruxure Building Operation (EBO) was implemented across the campus. Sensors from multiple buildings are able to be monitored and controlled through a single platform. The system in University of Nottingham uses an open, end-to-end IP architecture which enables fast connectivity of IoT devices. To reduce energy consumption, SmartX IP controllers and SmartX Living Space sensors were deployed. From continuous monitoring and control, 5% energy consumption reduction, 3% reduction on overall energy costs and 25% reduction of maintenance costs have been achieved. Table 7 shows a BEMS implementation across commercial and academic buildings.

4.2 Key Lesson Learned

The implementation of building and energy management systems in commercial buildings, airports and academic buildings highlighted several key lessons in scalability, stakeholder engagement and return of investment. Fig. 5 shows BEMS implementation across sectors.

The adaption of an open-architecture and open-communication protocol system by MSP airport and University of British Columbia allows future expansion, technology upgrade, and integration with IoT. Beyond the conventional BEMS platforms, integration of IoT provides a new perspective on how BEMS can monitor, control and contribute to net-zero building. BEMS system that traditionally serves as platform to control mechanical and electrical equipment, will now be able to serve as intelligent and distributed networks for continuous learning, adaptive optimization and cross-system collaboration. The next generation of BEMS with IoT integration will be inherently scalable and will be able to incorporate additional numbers of sensors, devices and subsystems while maintaining performance. There are two types of scalabilities, vertical and horizontal scalability. Vertical scalability, where the capability of the system is increased without adding new sensors, while horizontal scalability is expanding the system by adding more IoT devices, gateways, servers and other types of resources [116]. In enhancing occupants' comfort and increasing operational efficiency, integration of BEMS with IoT leverages AI and predictive analytics to control HVAC, lighting and security [117]. The open-architecture system also allows better collaboration of the stakeholders, such as the facilities managers, engineers, contractors and end-users [118]. Facilities managers are able to make decisions based on real-time data, while engineers and contractors can schedule preventive maintenance programs and users can access meaningful energy data.

However, it is important to ensure robust network security in the implementation of BEMS in any buildings. As the system is heavily relying on interconnected sensors, controllers and cloud-based platforms, it will be more vulnerable to cyber intrusions. A compromised BEMS can lead to unauthorized access to buildings, manipulation of operating systems and data breaches.

A recent review by [18] shows that implementation of BEMS in commercial buildings resulted in 15-30% of energy savings translating into annual savings of $1.5 – 4.4 million, depending on the building size and operational characteristics. Similarly, [119] demonstrated a 4.4-year payback period with the deployment of BEMS in a convenience store, with advanced logic control on the HVAC and the refrigeration system. These findings demonstrate that BEMS implementation can reduce operational energy consumption and deliver significant cost savings, thus supporting the economic viability of BEMS adoption with quantifiable returns on investment.

Table 7 BEMS implementation across commercial and academic buildings.

| Name | Type (m²) | Controls | Reference |
|---|---|---|---|
| The Wings, Brussels, Belgium | Commercial building | Metasys building automation system, Johnson Controls | [110] |
| Vorum, Nabburg, Germany | Healthcare | ABB's KNX ClimaECO | [111] |
| ArchiCe, Hanza Tower, Poland | Offices | ABB's i-bus® KNX | [112] |
| Minneapolis-St. Paul International Airport, Minneapolis-Saint Paul, United States | Airport | Honeywell's Niagara Framework™, Honeywell | [113] |
| The University of British Columbia, Vancouver, Canada | University | Combination of Siemens, Delta/ESC and Johnson Controls Inc. (JCI) | [114] |
| The University of Nottingham, United Kingdom | University | Schneider Electric's EcoStruxure Building Operation (EBO) | [115] |

## 5 Emerging Trends and Future Directions

BEMS discourse has changed dramatically in recent years due to technological improvements and a rising dedication to sustainability. NZEBs are becoming more important as climate change concerns rise, advancing innovative energy-saving and renewable resource solutions. This transformation is led by IoT-driven technologies that enable real-time data analysis, automation, and user interaction in energy management. As industry evolves, AI, Blockchain, and Metaverse will transform energy efficiency and building management. This section examines sustainable BEMS trends and future directions, focusing on AI and data-driven optimization, RE integration, occupant-centric designs, support policies, digital twins, and metaverse applications. Advanced technology and user-centric tactics are utilized in these sectors to offer complete energy management solutions for modern buildings. Considering such improvements, carbon emissions can potentially be minimized while occupant comfort and building performance are improved.

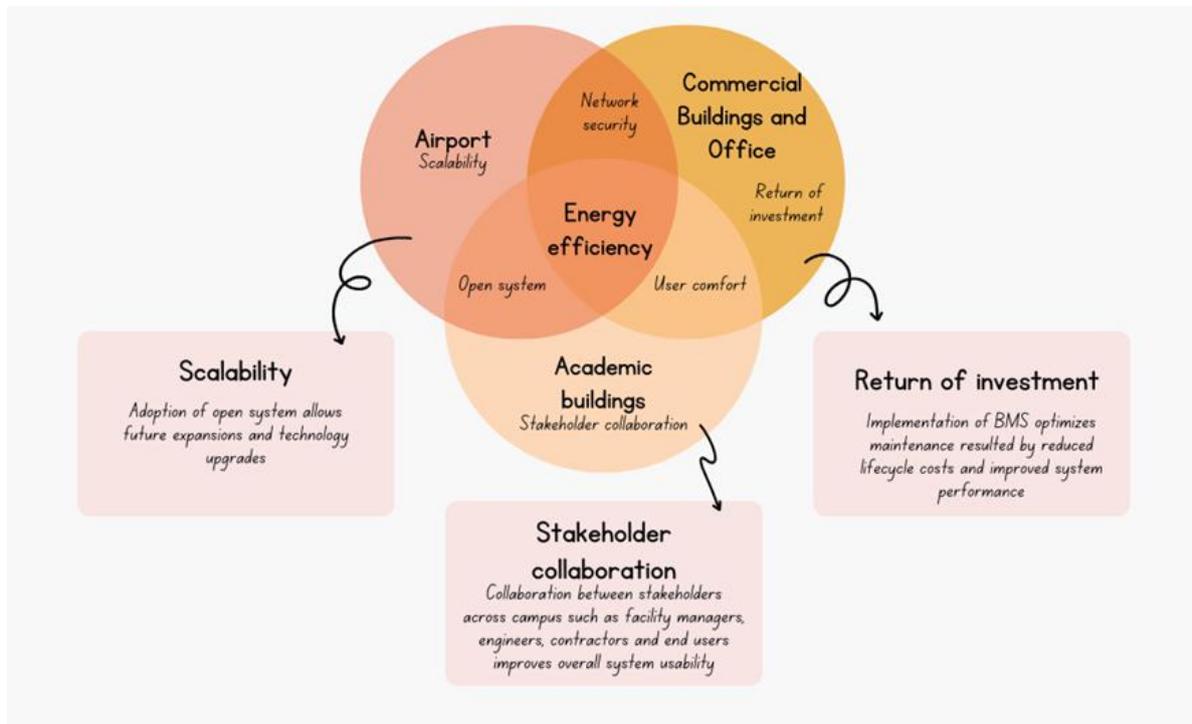

Fig. 5 BEMS implementation across sectors

5.1 AI and data Driven Optimization

The conventional building automation system is tedious and inconsistent, operating primarily in a one-way and manual system. The integration with IoT has significantly transformed these systems by providing comprehensive reporting capabilities which support improvement in decision making and energy saving initiatives. However, enhanced AI function and optimization in BEMS have increased the data usability and functionality in expanding system performance.

Data collected by sensor, meter and IoT application for real-time monitoring and control such as power consumption, temperature, humidity, light, motion, historical data and environmental for real-time monitoring and control, will undergo data preprocessing including filtering, normalization and imputation [80]. The adoption of AI and ML techniques in BEMS data is no longer a new thing where Artificial Neural Network (ANN), Deep Learning (DL) and Reinforcement Learning (RL) have applied for accurate energy forecasting, demand prediction and the optimization of HVAC system, lighting and RE scheduling [120]. ML and DL techniques have been applied in handling building energy (BE) system.

[121] has analyzed that hybrid and ensemble methods have good robustness performance in forecasting energy demand, consumption forecasting and load forecasting compared to traditional ML and Linear Regression (LR) methods. Furthermore, Model Predictive Control combined with data-driven methods has applied in improving control performance and reducing computational complexities during online implementation [122].

While current research demonstrates the effectiveness of AI and data driven optimization in enhancing BEMS, future trends are focused on developing more intelligent, autonomous and resilient systems. Explainable AI is becoming more important to increase transparency and interoperability of AI judgements for broader adoption and data reliability [123]. Furthermore, AI data driven involves developing hybrid AI models integrating IoT, blockchain, and edge computing for real-time and decentralized energy management [124]. As mentioned by [125], future research should focus on creating advanced optimization algorithms and sophisticated control designs to properly integrate RE resources and manage complicated building scenarios. The main goal is to shift towards autonomous and resilient AI-driven energy management system that can adapt to changing conditions and user needs, thus contributing to smarter, more sustainable and resilient built environments.

## 5.2 Integration with Renewables Energy

The integration of renewable energy sources into BEMS has become indispensable as the transition to sustainable energy practices accelerates. The implementation of PV systems, wind turbines, and other renewable technologies enables the generation of energy on-site, thereby reducing dependence on fossil fuels and enhancing grid resilience. Successful case studies, as shown in Table 2, demonstrate that effective renewable integration has resulted in substantial reductions in energy costs and greenhouse gas emissions, thereby demonstrating a successful alignment with global decarbonization objectives. Additionally, the optimization of energy consumption will be facilitated by the integration of renewable sources with energy storage technologies, such as batteries, which will enable buildings to store excess power for subsequent use during peak demand periods [126].

The seamless integration of renewable resources into building operations is facilitated by advanced energy management systems. For example, by employing IoT sensors and controllers, BEMS can intelligently regulate the energy generated from decentralized renewable sources, thereby guaranteeing that consumption is consistent with availability [127]. Furthermore, the efficacy of this integration is improved by the incorporation of energy forecasting models, which forecast energy generation based on historical performance data and meteorological conditions. Supporting energy independence and facilitating demand response capabilities, when necessary, energy management systems can coordinate with numerous renewable sources [128]. It is anticipated that the future of renewable integration in BEMS will capitalize on emergent technologies, such as blockchain, to facilitate decentralized energy trading, thereby improving the sustainability and resilience of energy systems [129]. This approach encourages the exchange of energy and optimizes the utilization of local resources by promoting peer-to-peer energy trading among buildings within a community. Such developments will enable the integration of renewables into BEMS, thereby facilitating the attainment of NZEBs and transforming urban energy ecosystems into more resilient and intelligent infrastructures.

## 5.3 Occupant-centric BEMS

Occupancy knowledge is essential for effective building management [130]. The development of BEMS is significantly influenced by occupant-centric design methodologies that incorporate performance metrics, including thermal comfort, energy efficiency, visual comfort, acoustic comfort, and wellbeing, as shown in Fig. 6. To be effective, BEMS must consider internal and external factors, such as stakeholders' requirements, budgets, and climatic conditions. This is because an improved occupancy experience can lead to increased productivity and satisfaction. BEMS can perpetually evolve toward more adaptive and resilient systems by incorporating these considerations into iterative design and modelling loops [131].

Recent technological advancements have enabled the development of more intricate methods for the collection and evaluation of occupant feedback, which are facilitated by simulation models and design applications [132]. This enables the dynamic modification of building environments based on real-time data. Intelligent technologies, including occupancy sensors, mood-based illumination, personalized climate controls, and AI-driven analytics, are creating more engaging environments that optimize energy performance and cater to individual preferences.

Additionally, research [129] suggests that individuals who experience a sense of tranquility in their environment tend to exhibit increased productivity and reduced tension. As a result, BEMS are evolving to prioritize the overall health, comfort, and productivity of occupants, in addition to energy optimization. This paradigm shift emphasizes the development of systems that reconcile technical efficiency with occupant-centric performance metrics. BEMS demonstrates efficacy in enhancing overall wellness by incorporating internal environmental quality (IEQ) elements, including air quality, thermal comfort, acoustic performance, and space planning [134]. The systems consistently adapt to the changing requirements of inhabitants by utilizing AI, parametric analysis, and uncertainty modelling to maintain optimal conditions.

In the future, occupant-centric BEMS will capitalize on interactive feedback mechanisms, customized user interfaces, and augmented reality to improve engagement [135]. BEMS are cultivating a culture of sustainability within the constructed environment by encouraging occupants to engage in energy-conserving practices, such as providing behavioral feedback or establishing personal comfort profiles. The development of smart buildings that are aligned with net-zero objectives and enhance urban quality of life will be dependent on the integration of building performance simulation, design applications, and real-time occupant interaction.

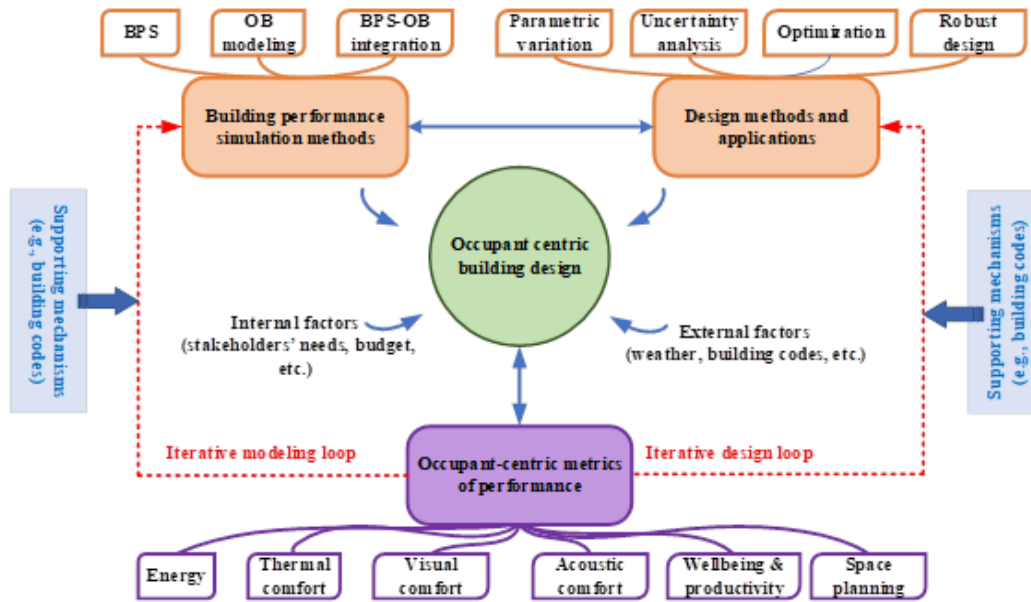

Fig. 6 Framework for Occupant-centric Building Design

5.4 Policy, Regulatory trend and Incentives Supporting NZEB and BEMS Deployment

The future landscape of the architectural industry and BEMS is increasingly shaped by emerging policies, regulatory frameworks, and financial incentives that align with global net-zero ambitions. The integration of technologies that support NZEB is being encouraged by policies as governments worldwide commit to ambitious climate targets, which is resulting in the increasing importance of sustainable construction practices. For instance, the Energy Performance of Buildings Directive (EPBD) of the European Union has implemented energy efficiency standards that encourage investment in advanced building technologies and innovation [136].

Retrofit Policy Instruments (RPIs) are also a component of potential policy advancements for energy-efficient buildings, with the objective of improving energy efficiency and reducing greenhouse gas emissions in existing residential structures [137]. These RPIs encompass financial incentives, research and services, assessment and disclosure tools, and direction and command instruments [138]. Building energy efficiency is also anticipated to be enhanced by enhanced standards and incentives for retrofitting HVAC systems [139]. Allied to this, the cultivation of renewable power and the decarbonization of the energy balance are acknowledged as essential strategies for reducing the energy consumption of buildings [140]. Building-related energy consumption has also been successfully decreased by local initiatives, including energy benchmarking, disclosure regulations, and mode-shift plans for urban transportation [141-142].

Financial incentives are essential for surmounting hurdles to the deployment of BEMS and energy-efficient enhancements. The financial feasibility of projects that are intended to achieve NZEB status is improved by mechanisms such as tax credits, grants, and subsidies, which reduce upfront costs. According to [143], targeted incentives can substantially increase the market demand for energy-efficient technologies. The implementation of building energy benchmarking policies, which necessitate proprietors to disclose their energy performance, can serve to enhance the market value of real estate and encourage further investment [142].

Simultaneously, policy developments are advancing toward more intelligent and flexible regulatory frameworks, such as model-predictive control, automated compliance monitoring, and semantic interoperability among complex systems [144]. The thermal building regulations are anticipated to be revised and extended in order to enhance the quality of building envelopes and minimize thermal losses [145]. Such developments demonstrate a paradigm shift from conventional compliance models to more proactive, technology-driven frameworks that improve both transparency and performance [146-147].

In the future, cooperation between researchers, industry stakeholders, and policymakers will be necessary to achieve the widespread deployment of BEMS and net-zero buildings. Accelerating the transition to carbon neutrality will require adaptive frameworks that change in tandem with technology, such as AI-based control systems, blockchain-enabled energy management, and clever retrofit techniques. The most viable route to promoting sustainable construction practices and reaching net-zero emission targets is a comprehensive strategy that incorporates financial incentives, retrofit plans, regulatory mandates, and renewable integration.

5.5 Digital Twins and Metaverse Applications for Building Energy Simulations

Emerging trends in IoT integration with BEMS are emphasizing the use of Digital Twin (DT) technology. DT is a dynamic, real-time virtual replicas of physical assets that use IoT sensor data for continuous monitoring, simulation and optimization to improve energy performance and operational decision making throughout the building lifecycle [148]. The effectiveness of DT and Metaverse applications in BEMS simulations relies on comprehensive and diverse data input to virtualize replicas the physical assets in enhancing energy performance and occupant comforts.

Research in DT has seen an exponential growth demonstrating strong scholarly and professional interest especially in the built environment. DT use AI algorithms to simulate, analyze, forecast energy performance, energy optimization, thermal management, optimum design, occupant well-being, building functionality, maintenance and modelling energy consumption by combining static BIM data with dynamic IoT sensor data [149]. Other than that, AI training engines have applied to generate datasets, and hybrid DT integration with Metaverse spaces has resulted in advanced scenario-based simulation for example fire, smoke and emergency egress as well as spatial behavior analysis through path and eye-tracking [150].

Currently, DT also applied district scale to emerging urban planning and helping to understand complex urban areas and optimizing energy systems [151]. The data-driven virtual representations of cities that combine real-time data, geographical information system (GIS) and AI analytics to support planning, simulation and energy monitoring. DT technology enables sustainable building energy management and cost reduction by monitoring, optimizing energy efficiency and predicting energy consumption in real time [152].

The future direction of DT and Metaverse application and building simulation is focused on highly integrated, intelligent and adaptable system. The key direction includes the development of standardized framework to guide DT implementations and focused on improving cybersecurity and implementing occupant behaviour models into DTs [153]. Future initiatives will focus on establishing interoperable and scalable framework for integrating heterogeneous systems into large infrastructure while maintaining robust data confidentiality and security [149].

Moreover, increasing international collaboration is required to advance and validate the scholarly research of DT as a solution of energy efficiency and thermal comfort [154]. [155] has mentioned the integration of ML techniques with DT technology is enhanced to increase simulation forecasting capacity and the obligatory of ethical governance into technological and policy framework as DTs grow more autonomous. Utilizing collaborative co-simulation platforms for urban DTs and developing innovative digital tools will assist to better assessing and evaluating energy use and demand, therefore simplifying the transition to more sustainable and energy-efficient built environments [156].

**6 Challenges and Research Gaps**

Despite the rapid growth of IoT and their increasing integration into BEMS, large scale deployment, effective systems remain constraints by multiple interrelated obstacles. Although IoT-enabled BEMS has shown great opportunity and strong potential in enhancing energy efficiency, operational intelligence and sustainability outcomes particularly in NZEB, several barriers impede their widespread adoption and long-term performance. The challenges were identified in several categories, there are technical challenges, financial challenges, organizational challenges and lastly the research gaps.

6.1 Technical Challenges

The implementation of IoT in BEMS encounters technical challenges related to hardware, software, data management and network integration. Most of IoT components and devices are self-battery and have a short service life, increasing maintenance costs and lowering system reliability [157]. Frequent battery replacement is not feasible, especially in BEMS applications where sensor downtime and data loss can occur when battery fails.

To develop smart and reliable BEMS systems that combine diverse systems, such as HVAC, lighting, meters, chilled water system, solar PV, and weather station, a strong network connectivity between the devices or the system is required. Each device provided by the vendor must have its own specification and characterization, creating a major challenge in integrating these heterogenous devices, as they often use different communication protocols and data formats. In practice, bridging technologies such as gateways, protocol translators, and data loggers are required to unify the IoT network. Semantic interoperability is necessary to link IoT devices, DT and older BMS platform in BEMS [81, 144] and without consistent standards, data consolidation and control across systems becomes a severe bottleneck.

NZEB operation requires prompt control to achieve maximum energy efficiency, minimize energy wastage while maintaining comfort, for example adjusting lighting and HVAC based on occupancy. IoT implementation in BEMS allows for real-time monitoring of energy consumption where the data generated by the sensors [158]. The biggest technical challenge is to ensure that BEMS can handle and respond to continuous IoT data in real-time. [132] stated there is an insufficiency of comprehensive energy efficiency strategies and real time implementation technologies in existing BEMS and the challenge arises due to several factors, for example communication error, or sensors failures.

Effective energy management in BEMS relies on accurate and reliability of the sensor data. Given that IoT sensors are frequently placed in hostile environments and are typically low-cost components prone to malfunctions, it is quite challenging to maintain their correct operation and prediction [159]. This requires a precise, automated, scalable, and flexible monitoring process, where the process is commonly known as sensor outlier detection to evaluate the behavior and performance within the IoT.

As IoT networks expand, security challenges inherently escalate, making devices more susceptible to unauthorized access and data breaches [77]. According to studies [19], IoT devices are hackable and each new device might present new security risks. To prevent disruptions, data theft, or malicious control of building systems, an IoT-enabled NZEB requires strong security at all tiers including secure firmware, encrypted connections and strict authentication).

Some NZEBs implemented BEMS with existing buildings that have conventional electrical wiring, communication network and structures. Integrating legacy BEMS with modern IoT platform presents several significant challenges, where these challenges are principally caused by the inherent characteristic of older systems, for example outdated protocols, limitations of software and hardware, and the requirement of the new, interconnected IoT environment [160]. This may lead to higher cost for replacement and integration and may affect maintenance efficiency and system scalability.

6.2 Financial Challenges

Transformation towards NZEBs led by IoT-driven technologies enables real-time monitoring, intelligent control, optimized energy performances and preventative maintenance implementation. However, implementation of smart buildings has faced challenges arising from various factors in terms of policy, financing and regulations. A study conducted by [161] identifies 23 barriers that limit the development of smart buildings, including lack of financial constraints, limitation on qualified professionals, government policy limitations, and regulatory barriers. While the technical potential of IoT is well recognized, financial constraints should be addressed to enable a large-scale adoption. The challenges center primarily on high upfront costs, difficulties in quantifying and predicting return on investment (ROI), and limitations in suitable financing models for software-driven and digital infrastructure.

One of the most highlighted barriers to IoT deployments to the built environment, whether it is a green field or the brown field is the substantial initial capital required. The high upfront costs are associated with expenditures on sensors, communication networks, data platforms, cybersecurity infrastructure and intensive integration with the existing building management systems. In a legacy building, costs of retrofitting the overall architecture includes additional wiring, structural upgrades, and interoperability solutions to enable communication between systems. Phased installation was usually required, which requires temporary shutdown of building services and extensive system testing, all of which further increase project duration, and will reflect in the higher project costs. In contrast, for new buildings, although the physical retrofitting is largely avoided, significant expenditure will be incurred during the design and construction phase to include installation of sensor networks, digital-ready HVAC and electrical systems, cybersecurity frameworks and integration with advanced building management system platforms. The high initial costs are consistently cited as the most significant obstacle for deployments, particularly when the funding mechanisms are limited or insufficient [162].

The high initial investment is accompanied by considerable uncertainty in the quantification of return on investment (ROI) due to lack of reliable and standardized methods in IoT-based systems. While implementation of IoT-based systems delivers a long-term benefit in terms of energy savings, reduced operating costs, improve fault detection and enhanced asset performance, a robust prediction of these gains at the pre-implementation stage remains highly uncertain. The financial returns are strongly influenced by factors such as occupant behaviors, weather variability, building schedules, fluctuations in energy tariffs, and data quality [163]. As a result, projected savings frequently deviated from the actual outcomes, introducing additional risks into investment evaluations. This uncertainty translates into extended payback periods, thus weakening the attractiveness of IoT investments under conventional budgeting criteria. Furthermore, advantages of IoT-based systems often come in a qualitative nature, such as improved indoor environmental quality, enhanced occupants' satisfaction, increased system reliability and proactive maintenance planning. These qualitative indicators are often not being captured within standard financial evaluation frameworks, thereby contributing to stakeholders' hesitation toward investing in an IoT based system [164].

IoT-based system adoption is also being restricted to limitations in financing structures and access to capital. Conventional financing models and loan instruments are typically designed for tangible assets with predictable depreciation and cashflows, whereas IoT investments are intangible. IoT-based systems are largely software-centric, rapidly evolving and tied to operational performance rather than direct revenue generation. Although emerging mechanisms such as energy performance contracts, leasing arrangements, and IoT-as-a-Service (IoTaaS) offer potential pathways to lower entry barriers by aligning payments with realized savings, their market penetration remains limited due to regulatory constraints, lack of standardized contract templates, and perceived investment risks [165]. In developing economies, the shortage of financing instruments designed for digital or green infrastructure, coupled with conservative credit markets, significantly constrains access to necessary capital.

6.3 Organizational Challenges

In contempt of advances in research and increasing pilot development in recent years, large scale deployment of NZEBs remains constrained by organizational and institutional challenges. Literature consistently indicates that beside technical and financial challenges, fragmented governance structures, skills shortages, misaligned incentives, and limited organizational readiness to manage complex, data-driven buildings contributes to slow NZEB adoption rate [167, 168, 174].

Fragmentation across governance, policy, and institutional frameworks: A recurring challenge is the fragmentation of NZEB governance across policy, regulatory, and market domains. Definitions of NZEB and net-zero carbon buildings vary widely across regions, jurisdictions, creating uncertainty for stakeholders and limiting knowledge transfer [174]. Supranational national and national ambitions often require translation into local building codes, enforcement mechanisms, and implementation guidance, resulting in a policy–practice gap that undermines effective deployment [167, 169].

Discontinuity across building lifecycle: NZEB delivery typically involves multiple actors across design, construction, commissioning, and operation, yet responsibilities for performance outcomes are rarely continuous across these phases. Studies highlighted that performance intent established during design could be lost during construction or operation due to organizational silos and weak handover processes [168, 170]. As a result, design knowledge or compliance does not reliably translate into sustained operational net-zero performance [27, 171].

Skills, knowledge, and workforce capacity gaps: The successful design, delivery and operation of NZEBs require specialized skills in integrated design, energy modelling, cross disciplinary knowledge, BEMS and performance monitoring. However, workforce shortages and limited training capacity remain widespread, particularly in HVAC operation, digital systems, and data interpretation [167]. Case studies from emerging-economy contexts further indicate reliance on imported expertise and limited local institutional learning, constraining scalability and long-term performance [27, 168].

Misaligned incentives, accountability, and risk allocation: Organizational incentives in the building sector often prioritize upfront capital cost, regulatory compliance, or certification achievement rather than post construction operational performance. Designers and contractors are rarely held accountable for post-occupancy outcomes, while building operators may lack both early design access, authority and resources to optimize NZEB operation [168]. In existing building to NZEB retrofit contexts, stakeholder resistance, uncertainty regarding benefits, and perceived operational risks further slow adoption [170]. These misalignments reinforce conservative decision-making and limit experimentation with advanced NZEB solutions [171].

Data governance, interoperability, and organizational readiness for digitalization: As NZEBs increasingly rely on IoT-enabled BEMS and data-driven operation, organizational challenges related to data governance and system integration have become more pronounced. The literature identifies unclear data ownership, interoperability limitations between proprietary systems, and cybersecurity concerns as significant barriers to effective operational coordination [172, 173]. Limited institutional readiness to manage large volumes of operational data further constrains the realization of potential performance gains [167, 170].

Regional inequities in organizational readiness: Organizational capacity to deliver NZEBs varies significantly across regions, contributing to uneven global adoption. Developed economies benefit from funding potential, stronger policy support, professional ecosystems, and demonstration projects, while emerging economies often face institutional capacity constraints, limited enforcement mechanisms, and climate-specific knowledge gaps [27, 174]. These disparities persist even where technical potential for NZEB deployment is high, underscoring the importance of institutional maturity and context-specific organizational support [167, 175].

6.4 Research Gaps

Despite extensive research on NZEBs several structural gaps remain that limit the application, reliability, and scalability of reported outcomes. While advances in IoT-enabled BEMS have expanded technical capabilities, existing studies remain fragmented across metrics, data sources, and evaluation approaches. The following gaps are consistently evident across the reviewed literature.

Lack of standardized performance metrics: Discussed in section 2.0, NZEB are defined and evaluated using heterogeneous metrics, including annual energy balance, carbon emissions, and energy flexibility; often more being framed using varying system boundaries and accounting assumptions. This inconsistency constrains comparison across studies and limits the transferability of findings. In parallel, research metrics are not consistently aligned with those used in regulatory frameworks and certification systems, further widening the gap between academic evaluation and practical implementation.

Limited real-world and longitudinal operational evidence: A substantial proportion of NZEB studies rely on simulation-based analysis or short-term monitoring, with limited examination of long-term operational performance under real occupancy conditions. Empirical evidence from living-lab studies further demonstrates that occupant behaviour, usage patterns, and operational uncertainty are dominant contributors to deviations between predicted and measured NZEB performance [176]. The scarcity of longitudinal operational datasets restricts understanding of performance persistence, system degradation, and the robustness of control strategies. A search in literature revealed studies based on extended BEMS monitoring in real operational settings, remains underrepresented but is necessary to address these limitations.

Insufficient integration of occupant centric, organizational and technical dimensions: Most NZEB optimization and BEMS studies prioritize technical performance. Reviews of smart building research show a growing shift toward human-cyber-physical system (HCPS) frameworks, yet human needs, roles, and decision-making remain weakly embedded in control logic and evaluation methodologies [177]. As organizational dimensions significantly influence operational outcomes, future research should more systematically integrate human-centric data streams, organizational workflows, and operational decision-making into IoT-enabled BEMS design and evaluation, moving beyond purely technology-driven optimization.

Fragmented evaluation frameworks for IoT-enabled BEMS: Although advanced IoT architectures and data-driven control strategies are increasingly reported, there is no widely accepted framework for evaluating their effectiveness within NZEBs. Existing studies often assess isolated performance indicators without considering trade-offs between energy performance, occupant comfort, operational complexity, and organizational feasibility. The absence of reproducible and transferable evaluation methodologies limits systematic comparison and evidence-based scaling of IoT-enabled BEMS solutions.

7 Conclusions

This article investigated the notion of IoT-driven BEMS as a key enabler for NZEBs. It emphasizes that achieving net-zero performances goes beyond energy-efficient design and renewable energy deployments to include sophisticated, data-driven operational control throughout the building lifecycle. Modern BEMS use high-resolutions sensing, real-time monitoring and automated decision-making to transform buildings into adaptive energy systems capable of balancing energy demand, occupant comfort, and energy reduction goals under dynamic operating conditions.

The study shows that effective NZEB implementation requires complete connectivity between building subsystems, RE technologies, energy storage and SG. BEMS can coordinate HVAC, lighting, distributed generation and DSM in real-time using IoT-enabled communication architectures, which are supported by standardized protocols and comprehensive data management frameworks. However, issues with heterogenous devices, data quality, cybersecurity, and system scalability continue to be significant impediments to wider implementation, resulting to performance gaps between simulated and real-world NZEB outcomes.

Looking forward, future BEMS development must prioritize scalable, secure, and interoperable systems based on high-quality real-world datasets and powerful AI approaches. The integration of predictive analytics, reinforcement learning, and DT technologies has the potential to significantly improve energy flexibility, fault resilience and long-term performance optimization. Continued collaboration among researchers, industry stakeholders and policymakers will be important in translating these advances into deployable solutions, allowing BEMS to play a key role in hastening the transition to net-zero buildings.

## Acknowledgement

The authors gratefully acknowledge the financial support from Sunway Education Group, Malaysia under project no STR-RES-SUST-NETZ-003-2025, which enabled the successful execution of this research. This work was further supported by the sustainability and research initiatives at Sunway University, particularly through the development and performance analysis of the University's Net Zero Carbon Building (Sunway EcoSphere) under the Faculty of Engineering and Technology. The EcoSphere initiative provided a living laboratory platform that facilitated real-world validation of low-carbon building strategies and advanced energy system integration. The authors also acknowledge the valuable contributions of faculty members, industry collaborators, and students whose technical expertise and support were instrumental to the completion of this study.

## Declaration of Competing Interest

The authors declare that they have no known competing financial interests or personal relationships that could have appeared to influence the work reported in this paper.

https://doi.org/10.1016/j.crsus.2024.100154

## Biographies

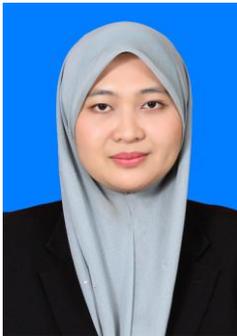

**Haizum Hanim Ab Halim** received BEng (Hons.) in Industrial Electronic Engineering at Universiti Malaysia Perlis (UniMAP), Malaysia in 2010, Master of Technical Education (Electrical Engineering) at Universiti Tun Hussein Onn Malaysia (UTHM), Malaysia in 2012 and PhD with Universiti Putra Malaysia (UPM), Malaysia. She is now working in Sunway University, Malaysia as a Post-Doctoral Research Fellow and actively involved in Net Zero Energy Annex Building, Sunway University. Her research interest includes Building Energy Management System, Internet of Things and Sustainability for Data Centre.

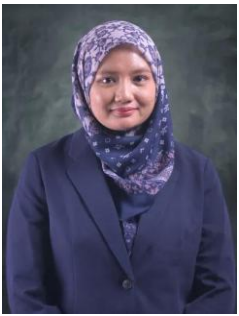

**Dalila Alias** received the B.Eng. degree from the Universiti Teknologi PETRONAS, in 2009 and the Ph.D. degree from the Universiti Putra Malaysia (Malaysia), in 2020. She is currently a postdoctoral research fellow at Sunway University. Currently, she plays an active role as a postdoctoral research fellow in the project management team for a university-led Net Zero building initiative, contributing to strategic planning, industry partnership, implementation and cross-disciplinary coordination aimed at achieving carbon neutrality.

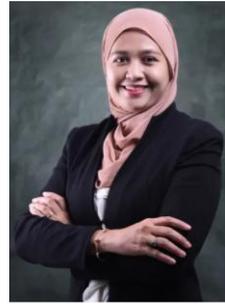

**Akmal Zaini Arsad** received the B.Sc (Hons) degree in physics, the M.Sc. degree in applied physics, and the Ph.D. degree in physics from the Universiti Kebangsaan Malaysia, Malaysia in 2011, 2012, and 2017, respectively. She has been a Postdoctoral Researcher with the School of Engineering, Faculty of Engineering and Technology, Sunway University since 2025. Her research interests focus on the integration of artificial intelligence in power and energy systems, particularly in hydrogen technologies, PV–battery storage, advanced control strategies, and building energy management systems. Her work encompasses both simulation-based analysis and real-world implementation solutions aimed at advancing building energy systems towards net-zero carbon targets.

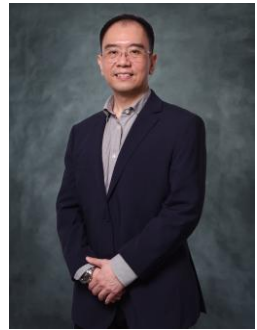

**Lewis Tee Jen Looi** received Bachelor of Science in Electrical Engineering (BSEE) at Purdue University, West Lafayette, USA, 2003; and Master of Science in Engineering (MSE) at Taylor's University, Lakeside Campus, Malaysia, 2019. He is a chartered engineer registered with Engineering Council United Kingdom (EC-UK), 2022 and a professional engineer registered with Board of Engineers Malaysia (BEM), 2023. He is working in Sunway University, Malaysia. His research interest includes net zero, sustainable energy, and decarbonisation strategies.

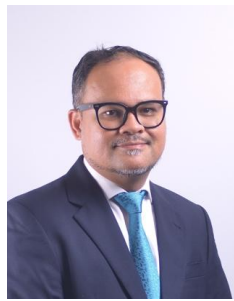

**Rosdiadee Nordin** received the B.Eng. degree from the Universiti Kebangsaan Malaysia, in 2001 and the Ph.D. degree from the University of Bristol, United Kingdom (U.K.), in 2011. He is currently a Co-Director and Professor at the Future Cities Research Institute (FCRI), a joint research partnership between Sunway University, Malaysia and Lancaster University, U.K. His research interests include Internet of Things wireless communications, edge-enabled sensing and AI-assisted wireless systems, with an emphasis on real-world deployments and scalable solutions for smart and sustainable cities.

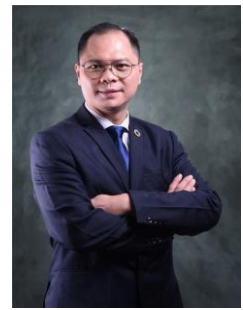

**Denny Ng Kok Sum** is the Dean of the Faculty of Engineering and Technology at Sunway University, Malaysia. He has authored over 260 publications and holds an h-index of 45. Actively involved in various professional bodies including IChemE and the Young Scientists Network–Academy of Sciences Malaysia (YSN-ASM), he is a Fellow of both IChemE (UK) and The Higher Education Academy (UK). He also holds professional registrations as a Chartered Engineer (Engineering Council UK), Professional Engineer (Board of Engineers Malaysia), and ASEAN Chartered Professional Engineer.